\documentclass[twocolumn]{pasj01} %proof
\usepackage{comment}  %
\usepackage{natbib}
\graphicspath{{figure/}}
\Received{2020 August 9}
\Accepted{2020 October 8}
\Published{$\langle$publication date$\rangle$}

\begin{document}

\title{Elemental Abundances of M Dwarfs Based on High-Resolution Near-Infrared Spectra: Verification by Binary Systems} %
\author{Hiroyuki Tako \textsc{Ishikawa}\altaffilmark{1,2,}$^{*}$,
Wako \textsc{Aoki}\altaffilmark{1,2},
Takayuki \textsc{Kotani}\altaffilmark{1,2,3},
Masayuki \textsc{Kuzuhara}\altaffilmark{2,3},
Masashi \textsc{Omiya}\altaffilmark{2,3},
Ansgar \textsc{Reiners}\altaffilmark{4}, Mathias \textsc{Zechmeister}\altaffilmark{4}
 }%
\altaffiltext{1}{Department of Astronomical Science,
The Graduate University for Advanced Studies, SOKENDAI,
 2-21-1 Osawa, Mitaka, Tokyo 181-8588, Japan}
 \altaffiltext{2}{National Astronomical Observatory of Japan, 2-21-1 Osawa, Mitaka, Tokyo 181-8588, Japan}
 \altaffiltext{3}{Astrobiology Center, 2-21-1 Osawa, Mitaka, Tokyo 181-8588, Japan}
 \altaffiltext{4}{Institut f\"{u}r Astrophysik, Georg-August-Universit\"{a}t, Friedrich-Hund-Platz 1, D-37077 G\"{o}ttingen, Germany}

\email{hiroyuki.ishikawa@grad.nao.ac.jp}%


\maketitle

\begin{abstract}
M dwarfs are prominent targets of planet search projects, and their chemical composition is crucial to understanding the formation process or interior of orbiting exoplanets. However, measurements of elemental abundances of M dwarfs have been limited due to difficulties in the analysis of their optical spectra. %
We conducted a detailed chemical analysis of five M dwarfs ($T_{\mathrm{eff}} \sim$3200--3800 K), which form binary systems with G/K-type stars, by performing a line-by-line analysis based on high-resolution ($R \sim$80,000) near-infrared (960--1710$\;$nm) spectra obtained with CARMENES. %
We determined the chemical abundances of eight elements (Na, Mg, K, Ca, Ti, Cr, Mn, and Fe), which are in agreement with those of the primary stars within measurement errors ($\sim$0.2 dex). %
Through the analysis process, we investigated the unique behavior of atomic lines in a cool atmosphere. %
Most atomic lines are sensitive to changes in abundance not only of the corresponding elements but also of other elements, especially dominant electron donors such as Na and Ca. %
The Ti I lines show a negative correlation with the overall metallicity at $T_{\mathrm{eff}} < $ 3400 K due to the consumption of neutral titanium by the formation of TiO molecules. %
These findings indicate that to correctly estimate the overall metallicity or the abundance of any element, we need to determine the abundances of other individual elements consistently. %
\end{abstract}

\section{Introduction} \label{sec:introduction}

M dwarfs are favorable targets for the discovery and detailed characterization of rocky planets, especially in the habitable zone, owing to their small mass, small size, and low luminosity compared to hotter Sun-like stars (e.g., \citealt{2016PhR...663....1S}).
One primary obstacle to the investigation of planets around M dwarfs is their faintness in the visible band. %
Through recent advancements in infrared detector technology, various infrared instruments have been developed that increase the feasibility of the study of M dwarfs because their peak fluxes are at infrared wavelengths.
Many current and future planet-search projects have set M dwarfs as the primary targets. %
For example, ground-based transit surveys such as MEarth (\citealt{2015csss...18..767I}), APACHE (\citealt{2013EPJWC..4703006S}), and SPECULOOS (\citealt{2018SPIE10700E..1ID}) mainly target M dwarfs.
Ground-based radial velocity surveys such as CARMENES (\citealt{2014SPIE.9147E..1FQ}), IRD (\citealt{2014SPIE.9147E..14K}), HPF (\citealt{2012SPIE.8446E..1SM}), and  SPIRou (\citealt{2014SPIE.9147E..15A}) also focus on M dwarfs.
Space-based surveys such as K2 (\citealt{2014PASP..126..398H}) have discovered a significant number of planets orbiting M dwarfs to date, and NASA's Transiting Exoplanet Survey Satellite (TESS; \citealt{2015JATIS...1a4003R}) is expected to detect approximately 500 planets around M dwarfs (\citealt{2018ApJS..239....2B}).

To further investigate the characteristics or habitability of planets found around M dwarfs, it is crucial to determine the detailed chemical compositions of the host M dwarfs.
Although the mean density of a rocky planet is estimated from its mass and radius, the internal structural features such as core size and mantle mineralogy, which control the surface habitability, cannot be determined without the constraints from the chemical composition (e.g., \citealt{2017A&A...597A..37D}, \citealt{2017ApJ...845...61U}).

A common measure of stellar chemical composition is the overall metallicity.
It is expressed as [M/H], which is the logarithm of the number ratio of all metals to hydrogen in the object of interest, normalized by the corresponding ratio in the Sun.
The iron abundance ratio [Fe/H], which is calculated in the same manner but with the number of Fe instead of all metals, is often used as a proxy for the overall metallicity.
The relation between the stellar metallicity and the planet occurrence rate has been discussed since \citet{2004A&A...415.1153S} and \citet{2005ApJ...622.1102F} established the case of Jupiter-mass planets around FGK stars.
\citet{2011ApJ...738...97B} and \citet{2012A&A...543A..89A} reported that the abundances of refractory elements such as Si, Mg, and Ti have a stronger correlation than the overall metallicity with the occurrence rate of giant planets.
These relations support the core-accretion scenario as the standard planet formation mechanism.
Such relations for M dwarfs have also been investigated, but they remain under discussion (e.g., \citealt{2018RMxAA..54...65H}, \citealt{2013A&A...551A..36N}). %
Further studies of M dwarfs with detailed abundances of refractory elements are needed to develop a better understanding of such trends. %

The chemical analysis of M dwarfs is hampered by their faintness and the complicated molecular absorption bands, which are difficult to model precisely and make it difficult to determine the continuum level and identify isolated absorption lines. %
Many previous studies have derived empirical relations between the overall metallicity and the metal-sensitive observable in photometry or low- to medium-resolution spectroscopy (e.g., \citealt{2005A&A...442..635B}, \citealt{2012ApJ...748...93R}, \citealt{2013AJ....145...52M, 2014AJ....147..160M}).
These relations are empirically calibrated by binary systems
consisting of FGK primaries and M-dwarf secondaries.
Stars in a binary system are formed together from the same molecular cloud; thus, their chemical abundances can be regarded as identical, typically within the level of 0.02 dex, and occasionally differing up to approximately 0.1 dex. %
This has been confirmed by several studies using binary systems comprised of FGK stars (e.g., \citealt{2004A&A...420..683D, 2006A&A...454..581D}, \citealt{2020MNRAS.492.1164H}).

The empirical calibrations above are useful for the exploration of global trends of overall metallicity. %
They do not, however, address abundance variations between the individual elements.
The most preferable way to directly address these abundances is to analyze individual absorption lines in high-resolution spectra with model atmospheres. %

In the last decade, the increasing number of available high-resolution near-infrared spectra of M dwarfs has led to advances in their chemical analysis.
Their near-infrared spectra are relatively free of the complicated and crowded molecular bands found in the visible spectra, and are therefore more suitable for line-by-line analysis of atomic lines.
\citet{2012A&A...542A..33O}, \citet{2016A&A...586A.100L}, and \citet{2017A&A...604A..97L} analyzed the atomic lines of various elements in high-resolution ($R \sim 50,000$) $J$-band CRIRES (\citealt{2004SPIE.5492.1218K}) spectra of early- to mid-M dwarfs to derive their metallicities by fitting synthetic model spectra.
They verified the reliability of the results using five M dwarfs in binary systems with FGK primaries.
They reported that the results of both components are generally consistent within 0.01--0.04 dex and at most 0.11 dex. %
\citet{2018A&A...610A..19R} performed overall spectral fitting on the APOGEE spectra ($H$-band, $R \sim 22,500$; \citealt{2017AJ....154...94M}) of 45 early- to mid-M dwarfs to determine the stellar parameters, including the metallicity.
\citet{2018A&A...620A.180R} conducted spectral fitting on a variety of atomic lines and OH molecular lines in the high-resolution ($R \sim 80,000$--$100,000$) visible and near-infrared spectra of 292 early- to late-M dwarfs taken by CARMENES to infer the stellar parameters, including the metallicity.
\citet{2019A&A...627A.161P} also derived the stellar parameters for a similar set of objects from the CARMENES spectra with a higher signal-to-noise ratio (S/N) %
by carefully selecting a limited number of absorption lines that are exclusively sensitive to the parameters of interest. %
Although their derived metallicities show better agreement with previous estimates through calibrated methods, there remain some systematic trends. %
This results in a narrower metallicity distribution than the estimates for the same objects in the literature. %
It was also noted that the metallicity derived only from the near-infrared spectra tends to be higher than that derived only from the visible spectra.
\citet{2020ApJ...890..133S} performed spectral fitting of %
APOGEE spectra to determine the metallicities of 21 early- to mid-M dwarfs, 11 of which form binary systems with FGK primary stars.
They confirmed that their results for the 11 M dwarfs are consistent with the metallicities of the corresponding primaries with a mean difference of 0.04 dex. %

Among these attempts, %
systematic differences in derived metallicities between different works are found in some cases.
They used different sets of absorption lines associated with different species to infer the overall metallicity. %
One possible improvement to reliably determine metallicity is to consider the abundance ratios of individual elements. %
\citet{2014PASJ...66...98T}, \citet{2015PASJ...67...26T}, \citet{2016PASJ...68...13T}, and \citet{2016PASJ...68...84T} %
determined the carbon and oxygen abundances and even the carbon isotopic ratios for M dwarfs based on CO and $\mathrm{H_{2}O}$ lines in $K$-band spectra with a resolution of $\sim$20,000.
\citet{2016ApJ...828...95V} found that the relative abundance of carbon to oxygen changes the pseudo-continuum of visible to near-infrared spectra of M dwarfs, which affects the derivation of the metallicity, particularly for empirical calibration methods.
\citet{2017ApJ...851...26V} developed a semi-empirical approach to determine the abundances of Fe and Ti.
They performed high-resolution ($R \sim 25,000$) $Y$-band spectroscopy on 29 M dwarfs that form physical binaries with FGK stars.
They found a discrepancy between their own equivalent width (EW) measurements and those predicted from the model and Ti abundance of the primary stars.
They empirically determined a transformation formula for the observed EWs to obtain an agreement with those based on models. %
\citet{2017ApJ...835..239S, 2018ApJ...860L..15S} determined the abundances of 13 elements for two early-M dwarfs and eight elements for one mid-M dwarf using atomic and molecular lines in the APOGEE spectra. %
\citet{2020ApJ...890..133S} also constrained the carbon and oxygen abundances of 21 M dwarfs %
in the process of estimating their effective temperatures based on the $\mathrm{H_{2}O}$ and OH molecular lines.
However, none of these studies have verified the resulting abundances of individual elements by testing the consistency of the FGK+M binary pairs.

Chemical analysis of M dwarfs based on model atmospheres has shown promise as a straightforward way to determine elemental abundances. %
However, the validity of the results and the potential problems inherent to the cool objects must be confirmed. %
In this paper, we present the abundance determination of eight individual elements for two early-M and three mid-M dwarfs using high-resolution near-infrared spectra.
To examine the abundances obtained by our analysis procedure, we target M dwarfs that belong to binary systems with G- or K-type stars. %
By examining the effects of variations in elemental abundance ratios assumed in the analysis on the abundances derived for %
certain elements, we demonstrate the importance of determining individual elemental abundances consistently. %
The chemical composition of M dwarfs and its effects on the spectra cannot be accurately described by the overall metallicities scaled from the solar chemical composition alone. %

We introduce our targets and data in Section \ref{sec:target data}. We describe how we exploit them to derive the abundances of individual elements and errors in Section \ref{sec:analysis} and present the results in Section \ref{sec:results}. We discuss the newly found problems and the remedies in Section \ref{sec:discussion}. In Section \ref{sec:summary}, we summarize the paper.

\section{Target selection and data} \label{sec:target data}
We selected five M dwarfs that are reported to form common proper-motion binary systems with G- or K-type stars. %
The high-resolution near-infrared spectra observed by CARMENES (Calar Alto high-Resolution search for M dwarfs with Exo-earths with Near-infrared and optical \'{E}chelle Spectrographs) of four of the M dwarfs are obtained from the CARMENES GTO Data Archive\footnote{http://carmenes.cab.inta-csic.es/gto/jsp/reinersetal2018.jsp} (\citealt{2018A&A...612A..49R}). %
We observed another M dwarf (G$\:$102-4) %
with CARMENES in a separate program (PI: M. Kuzuhara, Proposal Number: H18-3.5-091, F19-3.5-091). %
We stacked the data of G$\:$102-4 taken in three separate nights: December 14, 2018; January 28, 2019; and March 14, 2019.
The total exposure times for each night were 1798 s, 1398 s, and 2196 s, respectively, but the data in December and January fell short of the expected S/N due to the weather conditions.
Basic information and data quality for each object are provided in Table \ref{tab:targets}.
The near-infrared channel of CARMENES covers the wavelength range of 960--1710 nm ($Y$-, $J$-, and $H$-bands) with a spectral resolution of $R \sim$ 80,000.
There are gaps in the wavelength coverage of CARMENES. %
These are mostly less than 15 nm and
are located between adjacent echelle orders or in the center of each echelle order.
The gap in the center of each order is induced by a small gap between the two detectors.

The CARMENES data are provided as the one-dimensional spectra with vacuum wavelengths. They were reduced by the CARACAL pipeline (e.g., \citealt{2014A&A...561A..59Z, 2016SPIE.9910E..0EC}).
We normalized the spectra by fitting cubic spline curves to the continuum level of each echelle order. This procedure was conducted with the interactive mode of PyRAF\footnote{PyRAF is a product of the Space Telescope Science Institute, which is operated by AURA for NASA.}.
Further adjustment of the continuum level close to each absorption line was also made in the process of EW measurement. %
We merged the echelle orders into one spectrum for each object.
The spectra were Doppler-shifted to the stellar rest frame based on the radial velocity estimated by cross-correlation with the model spectra calculated from atomic line data. %

\begin{table*}%
  \tbl{Basic information of the target M dwarfs (in order of $T_{\mathrm{eff}}$)}{%
  %\small
  \renewcommand{\arraystretch}{1.0}
  \scalebox{0.95}[1.0]{ %
  \begin{tabular}{p{15mm}cccccp{13.5mm}ccc}%
    \hline\noalign{\vskip3pt}
        \multicolumn{1}{c}{Name}  &  RA  &  Dec  &  SpT\footnotemark[$*$]  &  $T_{\mathrm{eff}}$  &  $\log{g}$  &  Name$\;$of  &  SpT\footnotemark[$*$]$\,$of  &  [Fe/H]$\;$of  &  S/N\footnotemark[$\dag$]  \\ %
        &  \multicolumn{2}{c}{J2000.0}  && (K) && Primary  &  Primary  &  Primary  &  (1000 nm)  \\  [2pt]
    \hline\noalign{\vskip3pt}
        HD$\:$233153  &  \timeform{05h41m30.73s}  &  \timeform{+53D29'23.3''}  &  M1.0V  &  3765$\pm$60   &  4.70$\pm$0.13  &  HD$\:$37394  &  K1V  &  0.04$\pm$0.02  &  158  \\
        HD$\:$154363B  &  \timeform{17h05m13.78s}  &  \timeform{-05D05'39.2''}  &  M1.5V  &  3658$\pm$67  &  4.79$\pm$0.21  &  HD$\:$154363A  &  K5V  &  $-$0.62$\pm$0.05  &  182        \\
        BX$\:$Cet  &  \timeform{02h36m15.27s}  &  \timeform{+06D52'17.9''}  &  M4.0V  &  3284$\pm$60   &  4.94$\pm$0.13  &  HD$\:$16160A  &  K0V  &  $-$0.20$\pm$0.02  &  122  \\
        G$\:$102-4  &  \timeform{05h28m56.51s}  &  \timeform{+12D31'53.5''}  &  M4.0V  &  3246$\pm$57  &  4.97$\pm$0.20  &  HD$\:$35956  &  G0V  &  $-$0.05$\pm$0.02  &  107  \\
        $\rho^{01}$ Cnc B  &  \timeform{08h52m40.86s}  &  \timeform{+28D18'58.8''}  &  M4.5V  &  3166$\pm$61  &  4.94$\pm$0.14  &  HD$\:$75732  &  G8V  &  0.29$\pm$0.04  &  56        \\
    \hline\noalign{\vskip1pt}
  \end{tabular}}
  }\label{tab:targets}
  \begin{tabnote}
\hangindent6pt\noindent
\hbox to6pt{\footnotemark[$*$]\hss}\unskip Spectral type \hbox to6pt{\footnotemark[$\dag$]\hss}\unskip Signal-to-noise ratio
\end{tabnote}
\end{table*}

\section{Abundance analysis} \label{sec:analysis}
Our abundance analysis is based on the standard technique using a comparison between the EWs from observed spectra and those from synthetic spectra calculated with the one-dimensional LTE radiative transfer code.
The process is well established for visible spectra of FGK-type stars (e.g., \citealt{2019ARA&A..57..571J}).
However, there are difficulties in the analysis of near-infrared spectra of M dwarfs that have not yet been investigated well (see Section \ref{sec:discussion}). %
We carefully selected useful absorption lines for abundance analysis and employed valid atmospheric models for the temperature range of our targets.

\subsection{Stellar parameters} \label{sec:stellar_parameters}
For the calculation of model spectra, we need to set the stellar parameters, i.e., effective temperature $T_{\mathrm{eff}}$, surface gravity $\log{g}$, micro-turbulent velocity $\xi$, and elemental abundance ratios [X/H]. %

We adopted the literature values for $T_{\mathrm{eff}}$ and $\log{g}$ of the target and considered the errors propagated from the uncertainties reported.
The surface gravity is not directly reported in some cases, but can be calculated from the masses and radii. We referred to \citet{2015ApJ...804...64M} (hereafter Ma15) for the effective temperatures, masses, and radii of HD$\:$233153, BX$\:$Cet, and $\rho^{01}$ Cnc B. They derived the effective temperature by fitting the model spectra from the PHOENIX BT-Settl model atmosphere (\citealt{2012RSPTA.370.2765A}) to the observed medium-resolution ($R \sim$ 1000) visible spectra. They estimated the masses using the empirical mass--luminosity (mass--$M_K$) relation in \citet{2000A&A...364..217D} and the radii from $T_{\mathrm{eff}}$, bolometric flux $F_{\mathrm{bol}}$, and distance using the Stefan–Boltzmann law. $F_{\mathrm{bol}}$ was calculated by integrating the medium-resolution visible and near-infrared spectra complemented by the best-fit spectra from the BT-Settl model. We referred to \citet{2014MNRAS.443.2561G} for the same set of parameters of HD$\:$154363B. %
They derived $T_{\mathrm{eff}}$ and the mass in the same manner as Ma15 but employed the empirical $T_{\mathrm{eff}}$--radius relation in \citet{2013ApJ...779..188M} for the radius. We referred to \citet{2015ApJS..220...16T} for $T_{\mathrm{eff}}$ of G$\:$102-4, which was estimated using the $\mathrm{H_{2}O}$ indices in the $K$-band defined in \citet{2013ApJ...779..188M}. We used this $T_{\mathrm{eff}}$ to calculate the radius and mass by the radius--$T_{\mathrm{eff}}$ relation in Ma15 and the mass--radius relation in \citet{2019A&A...625A..68S}, respectively. Finally, we used the masses and radii above to calculate the surface gravity $\log{g}$ for each object.

The micro-turbulent velocity $\xi$ of M dwarfs can be assumed to be less than 1 km s$^{-1}$, as demonstrated observationally by \citet{2006ApJ...652.1604B} and theoretically by \citet{2009A&A...508.1429W}. We set $\xi$ to 0.5 km s$^{-1}$ for all objects for simplicity.
The changes in the resulting abundances by varying $\xi$ from 0.0 km s$^{-1}$ to 1.0 km s$^{-1}$ were included in the error estimate.

The $T_{\mathrm{eff}}$ and $\log{g}$ adopted in our analysis, and the associated errors, are given in Table \ref{tab:targets}. Note that we give the metallicities of the primaries in the table only for comparison purposes. %

\subsection{Synthetic spectra} \label{sec:model_atmosphere} %
We calculated the synthetic spectra with the one-dimensional LTE radiative transfer code, which is based on the same assumptions as the model atmosphere program of \citet{1978A&A....62...29T}, assuming individual elemental abundances of arbitrary values. %
The spectral line data were taken from the Vienna Atomic Line Database (VALD; \citealt{1999A&AS..138..119K}, \citealt{2015PhyS...90e4005R}).
Note that the spectral resolution does not affect the EW. %
For the atmospheric layer structure, we employed the one-dimensional plane-parallel LTE model atmosphere of MARCS\footnote{https://marcs.astro.uu.se/} (\citealt{2008A&A...486..951G}).
The grids of the atmospheric structure for $T_{\mathrm{eff}}$, $\log{g}$, $\xi$, and [Fe/H] are separated at intervals of 100 K, 0.5 dex, 1 km s$^{-1}$, and 0.25 dex, respectively. %
We interpolated the grid for our analysis. %

The targeted M dwarfs are hotter than 3000 K. %
In this $T_{\mathrm{eff}}$ range, dust formation in the atmosphere is negligible (e.g., \citealt{2013MmSAI..84.1053A}). %
We compared the temperature profiles as a function of pressure (P--T profiles) of the MARCS with those of the PHOENIX-ACES model for cool dwarfs presented by \citet{2013A&A...553A...6H}. %
The differences in temperature in the typical depth of line formation are less than 100 K, and thus do not have a significant impact on our results. %

\subsection{Spectral lines} \label{sec:spectral_lines} %
We identified atomic absorption lines by comparing the observed spectra to the synthetic ones calculated with the line list from VALD. %
We also added the data of FeH molecular lines, which form the prominent absorption band head around 1000 nm, using the line list edited by Plez (\citet{2012A&A...542A..33O}) to avoid contamination of the FeH lines on atomic lines.
We identified many atomic lines in the wide wavelength range of $Y$-, $J$-, and $H$-bands, but they could not all be used for the analysis because of the following difficulties:
(1) The existence of many weak but unidentified absorption lines makes it impossible to robustly determine the continuum level;
(2) Blending of other absorption lines; %
(3) Strong wing components appear even in relatively shallow absorption lines in the spectra of M dwarfs compared to those of FGK stars due to the high pressure of M-dwarf atmospheres (pressure broadening; see below for details); %
(4) The line strength is sensitive to changes in abundance of some elements other than the species responsible for the absorption line (see Section \ref{sec:insensitivity}).

Comparing the observed and synthetic spectra, we carefully selected the candidate absorption lines that were not affected significantly by issues (1) or (2). %
We made the final line list considering issues (3) and (4), and the number of candidate lines for each element.
For example, we excluded five Fe I lines from the analysis of mid-M dwarfs due to issues (3) and (4), whereas we did not exclude the Mg I line, which is a unique line available for the analysis, even though it had widely broadened wings and low sensitivity to Mg abundance. %
The information on the spectral lines used for abundance determination is presented in Table \ref{tab:linelist}.
Some lines of interest, such as the only Si I line unaffected by the telluric contamination, fall into the gaps of the CARMENES spectra and cannot be analyzed. %
\begin{table}%
  \tbl{Line list}{%
  \scalebox{1.0}[1.0]{ %{1.0}[0.91]
  \begin{tabular}{lcccc}%
    \hline\noalign{\vskip2.3pt} %
        Species & $\lambda$ (nm)\footnotemark[$*$] & $E_\mathrm{low}$ (eV)\footnotemark[$\dag$] & $\log gf$\footnotemark[$\ddag$] & vdW $\:$\footnotemark[${\S}$] \\[2pt] %
    \hline\noalign{\vskip3pt}
    Na I  &  1074.9393  &    3.191  &   $-$1.294 &    N/A  \\
    Na I  &  1083.7814  &    3.617  &   $-$0.503 &    N/A  \\
    Na I  &  1268.2639  &    3.617  &   $-$0.043 &   $-$6.653  \\
    Mg I  &  1183.1409  &    4.346  &   $-$0.333 &   $-$7.192  \\
    K I  &  1177.2861  &    1.617  &   $-$0.450 &   $-$7.326  \\
    K I  &  1177.6061  &    1.617  &    0.510 &   $-$7.326  \\
    K I  &  1243.5675  &    1.610  &   $-$0.439 &   $-$7.022  \\
    Ca I  &  1034.6654  &    2.932  &   $-$0.300 &   $-$7.480  \\
    Ca I  &  1195.9227  &    4.131  &   $-$0.849 &   $-$7.300  \\
    Ca I  &  1281.9551  &    3.910  &   $-$0.765 &   $-$7.520  \\
    Ca I  &  1282.7375  &    3.910  &   $-$0.997 &   $-$7.520  \\
    Ca I  &  1283.0568  &    3.910  &   $-$1.478 &   $-$7.520  \\
    Ca I  &  1291.2601  &    4.430  &   $-$0.224 &   $-$7.710  \\
    Ca I  &  1303.7119  &    4.441  &   $-$0.064 &   $-$7.710  \\
    Ca I  &  1306.1457  &    4.441  &   $-$1.092 &   $-$7.710  \\
    Ti I  &   967.8197  &    0.836  &   $-$0.804 &   $-$7.800  \\
    Ti I  &   969.1530  &    0.813  &   $-$1.610 &   $-$7.800  \\
    Ti I  &   972.1626  &    1.503  &   $-$1.181 &   $-$7.780  \\
    Ti I  &   973.1075  &    0.818  &   $-$1.206 &   $-$7.800  \\
    Ti I  &   974.6277  &    0.813  &   $-$1.306 &   $-$7.800  \\
    Ti I  &   977.2980  &    0.848  &   $-$1.581 &   $-$7.800  \\
    Ti I  &   979.0372  &    0.826  &   $-$1.444 &   $-$7.800  \\
    Ti I  &   983.4836  &    1.887  &   $-$1.130 &   $-$7.634  \\
    Ti I  &  1000.5831  &    2.160  &   $-$1.210 &   $-$7.780  \\
    Ti I  &  1005.1583  &    1.443  &   $-$1.930 &   $-$7.780  \\
    Ti I  &  1039.9651  &    0.848  &   $-$1.539 &   $-$7.810  \\
    Ti I  &  1058.7533  &    0.826  &   $-$1.775 &   $-$7.810  \\
    Ti I  &  1066.4544  &    0.818  &   $-$1.915 &   $-$7.810  \\
    Ti I  &  1077.7818  &    0.818  &   $-$2.666 &   $-$7.810  \\
    Ti I  &  1178.3767  &    1.443  &   $-$2.170 &   $-$7.790  \\
    Ti I  &  1180.0415  &    1.430  &   $-$2.280 &   $-$7.790  \\
    Ti I  &  1189.6132  &    1.430  &   $-$1.730 &   $-$7.790  \\
    Ti I  &  1281.4983  &    2.160  &   $-$1.390 &   $-$7.750  \\
    Ti I  &  1282.5180  &    1.460  &   $-$1.190 &   $-$7.790  \\
    Ti I  &  1292.3433  &    2.154  &   $-$1.560 &   $-$7.750  \\
    Cr I  &  1080.4319  &    3.011  &   $-$1.562 &   $-$7.780  \\
    Cr I  &  1082.4625  &    3.013  &   $-$1.520 &   $-$7.780  \\
    Cr I  &  1291.3622  &    2.708  &   $-$1.779 &   $-$7.800  \\
    Cr I  &  1294.0559  &    2.710  &   $-$1.896 &   $-$7.800  \\
    Mn I  &  1290.3289  &    2.114  &   $-$1.070 &    N/A  \\
    Mn I  &  1297.9459  &    2.888  &   $-$1.090 &    N/A  \\
    Fe I  &  1038.1843  &    2.223  &   $-$4.148 &   $-$7.800  \\
    Fe I  &  1039.8643  &    2.176  &   $-$3.393 &   $-$7.800  \\
    Fe I  &  1061.9630  &    3.267  &   $-$3.127 &   $-$7.780  \\
    Fe I  &  1078.6004  &    3.111  &   $-$2.567 &   $-$7.790  \\
    Fe I  &  1088.4739  &    2.845  &   $-$3.604 &   $-$7.810  \\
    Fe I  &  1089.9284  &    3.071  &   $-$2.694 &   $-$7.790  \\
    Fe I  &  1142.5447  &    2.198  &   $-$2.700 &   $-$7.820  \\
    Fe I  &  1161.0750  &    2.198  &   $-$2.009 &   $-$7.820  \\
    Fe I  &  1164.1446  &    2.176  &   $-$2.214 &   $-$7.820  \\
    Fe I  &  1178.6490  &    2.832  &   $-$1.574 &   $-$7.820  \\
    Fe I  &  1256.0432  &    2.279  &   $-$3.626 &   $-$7.820  \\
    Fe I  &  1288.3289  &    2.279  &   $-$3.458 &   $-$7.820  \\
        \hline\noalign{\vskip0pt} %
  \end{tabular}}
  } \label{tab:linelist} %
  \begin{tabnote}
\hangindent0pt\noindent %
\hbox to6pt{\footnotemark[$*$]\hss}\unskip%
 Wavelength in vacuum \quad
\hbox to6pt{\footnotemark[$\dag$]\hss}\unskip%
 Lower excitation potential \quad
\hbox to6pt{\footnotemark[$\ddag$]\hss}\unskip%
 Oscillator strength \quad
\hbox to6pt{\footnotemark[${\S}$]\hss}\unskip%
 Van der Waals damping parameter
\end{tabnote}
\end{table}

\begin{figure*}
 \begin{center}
  \includegraphics[width=160mm]{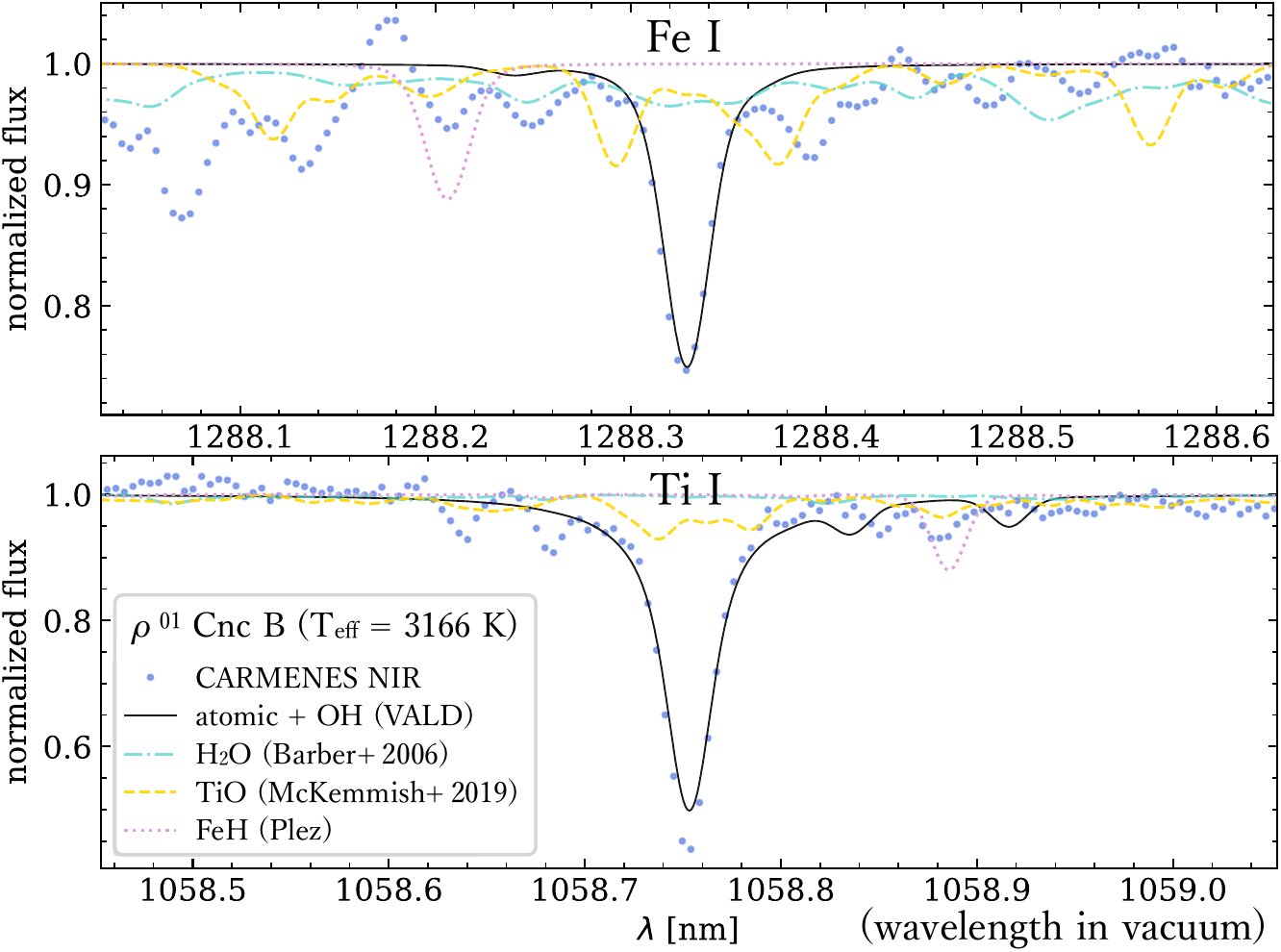} %
 \end{center}
 \caption{%
 Two atomic lines used in the analysis of $\rho^{01}\,$Cnc\,B ($T_{\mathrm{eff}}$ = 3166 K) and possible molecular lines in the vicinity.
 The solid-line synthetic spectrum (black line) is based on the stellar parameters and calculated using the line list from VALD, which includes atomic lines and OH lines.
 The other synthetic spectra are calculated with the $\mathrm{H_{2}O}$ line list of \citet{2006MNRAS.368.1087B} (cyan dot-dashed line),
 the TiO line list of \citet{2019MNRAS.488.2836M} (yellow dashed line),
 and the FeH line list of Plez (pink dotted line).
 The observed spectrum of $\rho^{01}\,$Cnc\,B is shown by blue dots. %
 }\label{fig:H2O+obj}
\end{figure*} %

Owing to the large surface gravity, the pressure broadening induced by neutral hydrogen and molecular hydrogen works strongly in M dwarfs. %
The pressure broadening dominates the total width of many lines and cannot be ignored, even in relatively weak lines.
We treated the broadening with the van der Waals (vdW) damping parameters calculated by \citet{2000A&AS..142..467B} if the data were incorporated in the VALD line lists.
For the absorption lines of Na and Mg without the reported damping data, we adopted the hydrogenic approximation of \citet{1955psmb.book.....U} with an enhancement factor, which is introduced conventionally to model the wings correctly.
The enhancement factors applied in previous studies are mostly between 1.0 and 5.0. %
We found that an enhancement factor of 1.0--1.9 closely reproduces the abundance results for the other elements obtained by the analysis using the vdW damping parameters of \citet{2000A&AS..142..467B}. %
We adopted 1.4 for Mn based on the empirical approximation by \citet{1998A&A...331.1051R}, and 1.0 for Na because \citet{1955ZA.....36..101W} demonstrated that the lines of alkali metals can be reproduced well without an enhancement factor.
We confirmed that these values are adequate for visual comparison between the line profiles of the models and observed data. %
When we varied the enhancement factor within 1.0--1.9, the variations in the resulting abundances from those lines were at most 0.05 dex.

We ascertained that the contamination of stellar $\mathrm{H_{2}O}$ lines is negligible in the wavelength region used in the present work, although the $K$-band spectra and some parts of the $H$-band spectra of M dwarfs were reported to be affected by numerous weak lines of $\mathrm{H_{2}O}$ changing the apparent continuum level (\citealt{2015PASJ...67...26T}). %
Examples of the small contribution of $\mathrm{H_{2}O}$ lines around the atomic lines used for mid-M dwarfs are shown in Fig. \ref{fig:H2O+obj}. %
For each panel, approximately 300 $\mathrm{H_{2}O}$ lines within 0.6 nm of the atomic line are included in the calculation. %
In the spectra of mid-M dwarfs, the depths of the most $\mathrm{H_{2}O}$ lines are significantly less than 1\,\%
 (and at most $\sim$5\,\%) of the continuum level, and they are sufficiently sparse not to affect the continuum flux drastically.
The contribution of $\mathrm{H_{2}O}$ is even smaller in the spectra of early-M dwarfs.

The possible contributions of the TiO and FeH lines are also shown in Fig. \ref{fig:H2O+obj}.
Most of the relatively strong (depth $\sim$10\,\%) TiO lines seen in the synthetic spectra were not found in the observed data.
The data of TiO lines in the near-infrared wavelength range are not sufficiently accurate for use in the analysis of high-resolution spectra, although we employed the up-to-date TiO line list, Toto (\citealt{2019MNRAS.488.2836M}).
The figure provides a conservative estimate indicating that the possible molecular lines do not have a significant effect on our analysis of atomic lines.

\subsection{Equivalent width measurement}\label{sec:ew_measurement}
We measured the EWs (e.g., \citealt{2018A&ARv..26....6N}) of the absorption lines using the PyRAF task ``splot’’ by fitting Gaussian profiles to the observed line profiles. For strong lines with wide wings that cannot be reproduced sufficiently by Gaussian profiles, we fitted Voigt profiles instead.
For the apparently overlapping absorption lines, we simultaneously fitted multiple Gaussian and/or Voigt profiles using the deblend fitting mode of splot. %
In the process of profile fitting, we redetermined the continuum level by visually checking the spectral range of more than $\sim$2 nm around the absorption line. %

\begin{figure}
 \begin{center}
  \includegraphics[width=80mm]{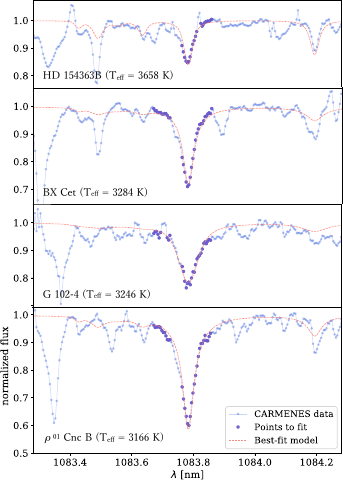}
 \end{center}
 \caption{CARMENES spectra (blue and purple dots) and synthetic model spectra calculated with the best-fit abundances (red dashed lines) around the Na I line at 1083.78 nm.
 The larger purple dots are the data used for the fitting (1083.67--1083.86 nm). %
 The wavelength range blended with unidentified lines (1083.73--1083.76 nm) and some outliers were excluded from the fitting.
 }\label{fig:Na_fitting}  %
\end{figure}

Although sodium is one of the most important elements in the temperature regime of M dwarfs (see Section \ref{sec:insensitivity} for details), there are no absorption lines suitable for the above procedure for EW measurement.
The self-blending of multiple Na I lines, the heavy blending of other spectral lines, or the overly broad wings deviated from the Gaussian or Voigt profiles hindered measurement with splot.
Instead, we determined the EWs of Na I lines with model fitting, treating the Na abundances and Gaussian width (representing the spectral resolution and macro-turbulence) as free parameters. %
An example of the Na I line at 1083.78 nm is shown in Fig. \ref{fig:Na_fitting}. This apparently single line actually consists of three lines, which are simultaneously used to fit the model spectra. %
The wavelengths used in the fitting are selected to cover from the line wings to the core while excluding the blending of unidentified absorption lines, telluric contamination, significant asymmetry, and outliers.
Note that the exclusion of the outliers does not significantly change the final result.

\begin{figure}
 \begin{center}
  \includegraphics[width=80mm]{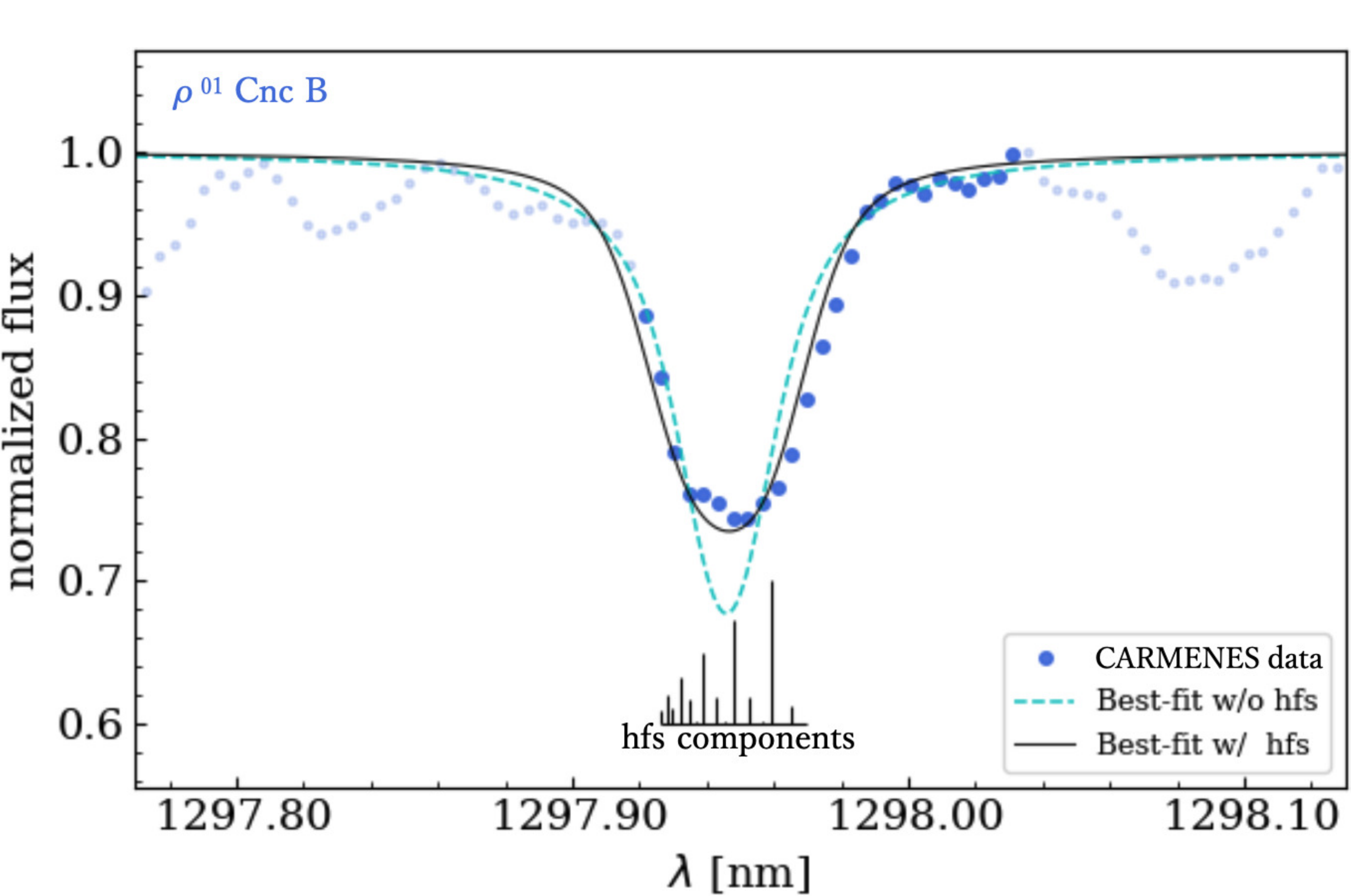}
 \end{center}
 \caption{
 Comparison between the best-fit model spectra calculated with and without considering the hfs splitting of the Mn I line at 1297.9 nm (black solid line and cyan dashed line, respectively).
 The blue dots are the CARMENES spectra of $\rho^{01}\,$Cnc\,B used for the fitting (the small pale dots are not used because of the line blending or uncertain continuum level).
 The positions of the hfs components are represented by vertical lines, the lengths of which are proportional to the gf-values. %
 }\label{fig:Mn_fitting}  %
\end{figure}

We also applied the fitting of model spectra to the manganese lines to consider the hyperfine structure (hfs) splitting. %
We employed the experimental oscillator strengths (gf-values) of individual hfs components reported by \citet{2011A&A...525A..44B} %
to reproduce the broad line profile of the Mn I lines
that could not be fitted satisfactorily by model spectra, calculated by assuming it to be a single line. %
Figure \ref{fig:Mn_fitting} shows the spectral fitting of the Mn I line at 1297.95 nm of $\rho^{01}\,$Cnc\,B with the contribution of each hfs component illustrated.

For G$\:$102-4, the line profiles are generally broad and the best-fit value of the Gaussian width is approximately 12 km s$^{-1}$, although the other objects indicate a width of $\sim$5 km s$^{-1}$.
This line broadening suggests a large rotational velocity. However, the line profile is well approximated by a Gaussian profile in the quality of our spectra; hence, we did not apply a more detailed rotational profile fitting in the present work.
We also found that G$\:$102-4 exhibits H$\alpha$ emission in the visible data of CARMENES, which indicates high magnetic activity and supports the possibility of rapid rotation.  %

All the measured EWs are listed in Table \ref{tab:EWs_excerpt}.
\begin{table*}%
  \tbl{Measured EWs in units of [pm] \footnotemark[$*$] %
  }{%
  \begin{tabular}{lccccccc}%
    \hline\noalign{\vskip3pt} %
Species & $\lambda$ (nm)\footnotemark[$\dag$] & HD$\:$233153 & HD$\:$154363B & BX$\:$Cet & G$\:$102-4 & $\rho^{01}\,$Cnc\,B \\[2pt]
\hline\noalign{\vskip3pt}
Na I  &  1074.9393  &    0.00  &    2.27  &    0.00  &    0.00  &    0.00  \\
Na I  &  1083.7814  &    0.00  &    4.99  &   14.14  &   16.15  &   25.08  \\
Na I  &  1268.2639  &   30.97  &    0.00  &    0.00  &    0.00  &    0.00  \\
Mg I  &  1183.1409  &   98.28  &  106.90  &   57.98  &   65.08  &   52.86  \\
K I  &  1177.2861  &   17.52  &   19.71  &    0.00  &   50.90  &    0.00  \\
K I  &  1177.6061  &    0.00  &    0.00  &    0.00  &  119.30  &    0.00  \\
K I  &  1243.5675  &   27.63  &   24.99  &   51.23  &    0.00  &    0.00  \\
Ca I  &  1034.6654  &  103.70  &   82.63  &    0.00  &    0.00  &    0.00  \\
Ca I  &  1195.9227  &    0.00  &    6.84  &    0.00  &    0.00  &    0.00  \\
Ca I  &  1281.9551  &   21.85  &    0.00  &   20.16  &    0.00  &   22.04  \\
        \hline\noalign{\vskip0pt} %
  \end{tabular}} \label{tab:EWs_excerpt}
  \begin{tabnote}
\hangindent0pt\noindent %
 \hbox to6pt{\footnotemark[$*$]\hss}\unskip%
An excerpt is shown here. The entire table is available online. \\ %
 \hbox to6pt{\footnotemark[$\dag$]\hss}\unskip%
 Wavelength in vacuum \\
\end{tabnote}
\end{table*}

\subsection{Abundance determination}\label{sec:abundance_determination}
We calculated the theoretical EWs from the synthetic spectra described in Section \ref{sec:model_atmosphere}. %

In the first step, we adopted a model atmosphere with the solar composition (\citealt{2009ARA&A..47..481A}).
We also assumed all elemental abundances to be solar values when initially calculating the synthetic spectra.
We varied the elemental abundance [X/H] in the spectral synthesis until the theoretical EW of the $i \,$th line of the element X matched the observed EW$_i$.
We determined the elemental abundance [X/H]$_i$ for each absorption line $i$, and obtained the abundance of the element X by averaging [X/H]$_i$ from all lines of X.

In the second step, we updated the model atmosphere by adopting the overall metallicity ([M/H]) that corresponds to the Fe abundance ratio ([Fe/H]) determined by the previous step.
In the spectral synthesis, we adopted the individual elemental abundances determined by the previous step as the assumed values. %
For the abundances of $\alpha$ elements that were not measured in the present work (e.g., C, O, Si, S), we adopted the average of the abundance ratios of measured $\alpha$ elements (e.g., Mg, Ca; we did not use Ti because of the large uncertainty discussed in Section \ref{sec:titanium}). %
In addition, we assumed the abundance ratios ([X/H]) of all the other unmeasured elements to be identical to that of iron ([Fe/H]). %
Based on these settings, we redetermined the abundances of individual elements in the same manner as in the first step.

Based on these results, we again updated the model atmosphere and elemental abundances assumed in the next step. %
These steps were iterated until the resulting [Fe/H] agreed with the [M/H] assumed in the model atmosphere within 0.005 dex to obtain the final results. %

\subsection{Error estimation} \label{sec:error_estimate}
We calculated the total error of the abundance for each element as the quadrature sum of four types of errors.
The error size from each source is tabulated in Table \ref{tab:result}, and the contribution of each to the total error is described in Section \ref{sec:error_result}.

(1) Errors due to random noise in spectral lines, uncertainties in the line data, and measurement uncertainty ($\sigma_\mathrm{SEM}$). %
These are estimated as the standard error of the mean (SEM) by dividing the standard deviation of the resulting [X/H] from individual absorption lines by the square root of the number of absorption lines used in the analysis ($N_\mathrm{line}$).
For the species for which we used fewer than four absorption lines, we substituted the standard deviation of Fe abundance and divided it by the square root of the number of lines of the species.
This is because the number of Fe I lines is at least four, with the exception of G$\:$102-4.
We used only two Fe I lines for G$\:$102-4, but the standard deviation from the two lines is comparable to the standard deviation from the results of five Fe I lines for another mid-M dwarf, $\rho^{01}\,$Cnc\,B.

(2) Errors propagated from uncertainties of the three stellar parameters: $T_{\mathrm{eff}}$, $\log{g}$, and $\xi$ ($\sigma_{T_{\mathrm{eff}}}$, $\sigma_{\log{g}}$, and $\sigma_{\xi}$, respectively). We performed the entire analysis procedure by assuming the minimum or maximum value in the uncertainty range for each parameter to obtain the change in the results as the error. %

(3) Errors caused by uncertainties in the abundances of other elements ($\sigma_\mathrm{OE}$).
Here, we conservatively adopted 0.2 dex as the uncertainties, referring to the errors due to the other sources (1), (2), and (4).
For each element, we repeated the entire procedure assuming the abundances of the other elements homogeneously to be 0.2 dex higher or lower than our final results.
We adopted the variation in the result as the error size.

(4) Errors arising from the uncertainty of the continuum-level determination due to noise or possible weak features around the spectral lines ($\sigma_\mathrm{cont}$).
We estimated the error by abundance analysis using the same procedure by changing the continuum level.
The possible ranges of the continuum level were estimated by visual inspection of the Fe I and Na I lines.
The rate of change in EW when the continuum level was varied within each of these ranges was found to correlate with the EW value of the absorption line.
To estimate the rate of change for all the other lines based on this correlation, we performed linear fitting to obtain the empirical equations as follows:
\begin{eqnarray}
  \Delta \mathrm{EW_{min}} / \mathrm{EW} & = & - 0.145 + 0.0093 \, \mathrm{EW}, \label{eq:er1} \\
  \Delta \mathrm{EW_{max}} / \mathrm{EW} & = & \:\ \ 0.230 -0.0105 \, \mathrm{EW}, \label{eq:er2} \\
  \Delta \mathrm{EW_{min}} / \mathrm{EW} & = & - 0.049 + 0.0001 \, \mathrm{EW}, \label{eq:er3} \\
  \Delta \mathrm{EW_{max}} / \mathrm{EW} & = & \:\ \ 0.157 -0.0042 \, \mathrm{EW}, \label{eq:er4}
\end{eqnarray}
where the EWs are measured in units of pm.
$\Delta \mathrm{EW_{min}}$ and $\Delta \mathrm{EW_{max}}$ are the changes in EWs when the minimum and maximum possible continuum levels are adopted, respectively.
Equations (\ref{eq:er1}) and (\ref{eq:er2}) are valid for relatively weak lines with EWs less than $\sim$10 pm, and
(\ref{eq:er3}) and (\ref{eq:er4}) correspond to stronger lines. %
We verified from the Fe I and Na I lines that the departure of the estimate by the empirical relation from that by the visual inspection for each line is almost less than 10\,\%, which has a negligible effect on the final error size.

We applied these equations to all absorption lines to estimate the errors associated with the continuum level determination. %
A $\Delta \mathrm{EW_{min}} / \mathrm{EW}$ or $\Delta \mathrm{EW_{max}} / \mathrm{EW}$ value of 0.05 was adopted in the case for which the values obtained from the above equations were smaller than 0.05.
We took the average of the errors arising from $\Delta \mathrm{EW_{min}}$ and $\Delta \mathrm{EW_{max}}$ as the error for each line.
We calculated the root mean square (RMS) of the errors for all lines of an element and divided it by the square root of the number of absorption lines to obtain the standard error of the resulting abundance of the element.
The rate of change in EW ranged within 5--20\,\%
 and the resulting abundance error was sufficiently small to be overwhelmed by other error sources.

\section{Results} \label{sec:results}
\begin{table*}[pbth]
  \tbl{Abundance results with the individual contribution of each error source}{%
  \scalebox{1.0}[1.0]{
  \begin{tabular}{llccccccccc}%
    \hline\noalign{\vskip3pt}
        \multicolumn{1}{c}{Object} & Element &  [X/H]  &  $N_\mathrm{line}$  &  $\sigma_\mathrm{SEM}$  &  $\sigma_{T_{\mathrm{eff}}}$  &  $\sigma_{\log{g}}$  &  $\sigma_{\xi}$  &  $\sigma_\mathrm{OE}$  &  $\sigma_\mathrm{cont}$  &  $\sigma_\mathrm{Total}$ \\
    \hline\hline\noalign{\vskip3pt}
        HD$\:$233153
        & Na &  0.06 &     1 &  0.20 &  0.00 &  0.02 &  0.00 &  0.02 &  0.03 &  0.20 \\
        & Mg &  0.09 &     1 &  0.20 &  0.08 &  0.02 &  0.01 &  0.09 &  0.05 &  0.24 \\
        & K &  0.15 &     2 &  0.14 &  0.02 &  0.04 &  0.02 &  0.05 &  0.04 &  0.16 \\
        & Ca &  0.22 &     7 &  0.08 &  0.01 &  0.02 &  0.01 &  0.05 &  0.03 &  0.10 \\
        & Ti &  0.36 &    12 &  0.04 &  0.04 &  0.01 &  0.04 &  0.14 &  0.02 &  0.16 \\
        & Cr &  0.23 &     3 &  0.11 &  0.04 &  0.01 &  0.03 &  0.08 &  0.06 &  0.16 \\
        & Mn &  0.31 &     2 &  0.14 &  0.05 &  0.01 &  0.02 &  0.10 &  0.04 &  0.18 \\
        & Fe &  0.14 &    10 &  0.06 &  0.04 &  0.01 &  0.03 &  0.09 &  0.03 &  0.13 \\
        \hline
        HD$\:$154363B
        & Na & $-$0.41 &     2 &  0.13 &  0.04 &  0.03 &  0.01 &  0.02 &  0.03 &  0.15 \\
        & Mg &  0.17 &     1 &  0.19 &  0.04 &  0.11 &  0.01 &  0.16 &  0.05 &  0.27 \\
        & K & $-$0.05 &     2 &  0.13 &  0.07 &  0.11 &  0.02 &  0.06 &  0.03 &  0.20 \\
        & Ca & $-$0.13 &     6 &  0.05 &  0.03 &  0.07 &  0.01 &  0.07 &  0.03 &  0.12 \\
        & Ti &  0.05 &    13 &  0.03 &  0.00 &  0.09 &  0.04 &  0.19 &  0.02 &  0.22 \\
        & Cr & $-$0.36 &     2 &  0.13 &  0.01 &  0.00 &  0.02 &  0.10 &  0.06 &  0.18 \\
        & Mn & $-$0.47 &     1 &  0.19 &  0.01 &  0.09 &  0.03 &  0.15 &  0.05 &  0.26 \\
        & Fe & $-$0.34 &     9 &  0.06 &  0.00 &  0.04 &  0.03 &  0.13 &  0.03 &  0.15 \\
        \hline
        BX$\:$Cet
        & Na & $-$0.10 &     1 &  0.02 &  0.08 &  0.07 &  0.00 &  0.08 &  0.03 &  0.14 \\
        & Mg & $-$0.17 &     1 &  0.02 &  0.04 &  0.11 &  0.01 &  0.19 &  0.04 &  0.23 \\
        & K & $-$0.18 &     1 &  0.02 &  0.16 &  0.14 &  0.01 &  0.10 &  0.04 &  0.24 \\
        & Ca & $-$0.14 &     3 &  0.01 &  0.06 &  0.07 &  0.01 &  0.13 &  0.03 &  0.16 \\
        & Ti &  0.00 &    14 &  0.03 &  0.10 &  0.13 &  0.02 &  0.27 &  0.02 &  0.32 \\
        & Cr & $-$0.26 &     2 &  0.02 &  0.04 &  0.03 &  0.01 &  0.12 &  0.06 &  0.15 \\
        & Mn & $-$0.08 &     1 &  0.02 &  0.06 &  0.07 &  0.01 &  0.16 &  0.06 &  0.20 \\
        & Fe & $-$0.21 &     4 &  0.01 &  0.05 &  0.05 &  0.02 &  0.14 &  0.04 &  0.16 \\
        \hline
        G$\:$102-4
        & Na & $-$0.04 &     1 &  0.14 &  0.08 &  0.12 &  0.00 &  0.09 &  0.04 &  0.22 \\
        & Mg & $-$0.02 &     1 &  0.14 &  0.04 &  0.18 &  0.01 &  0.20 &  0.04 &  0.31 \\
        & K & $-$0.03 &     2 &  0.10 &  0.18 &  0.26 &  0.01 &  0.13 &  0.03 &  0.36 \\
        & Ca & $-$0.26 &     2 &  0.10 &  0.03 &  0.08 &  0.00 &  0.11 &  0.04 &  0.18 \\
        & Ti &  0.03 &    13 &  0.05 &  0.10 &  0.23 &  0.02 &  0.28 &  0.02 &  0.38 \\
        & Cr & $-$0.31 &     1 &  0.14 &  0.03 &  0.04 &  0.01 &  0.12 &  0.08 &  0.21 \\
        & Mn & $-$0.17 &     2 &  0.10 &  0.08 &  0.16 &  0.02 &  0.18 &  0.04 &  0.28 \\
        & Fe & $-$0.18 &     2 &  0.10 &  0.05 &  0.09 &  0.02 &  0.14 &  0.06 &  0.21 \\
        \hline
        $\rho^{01}\,$Cnc\,B
        & Na &  0.42 &     1 &  0.13 &  0.14 &  0.19 &  0.02 &  0.16 &  0.03 &  0.31 \\
        & Mg &  0.23 &     1 &  0.13 &  0.08 &  0.20 &  0.02 &  0.17 &  0.04 &  0.30 \\
        & Ca &  0.23 &     5 &  0.08 &  0.08 &  0.15 &  0.02 &  0.14 &  0.03 &  0.24 \\
        & Ti &  0.59 &    13 &  0.03 &  0.17 &  0.31 &  0.06 &  0.31 &  0.02 &  0.48 \\
        & Cr &  0.14 &     1 &  0.13 &  0.06 &  0.09 &  0.02 &  0.12 &  0.09 &  0.22 \\
        & Mn &  0.43 &     1 &  0.13 &  0.11 &  0.18 &  0.04 &  0.17 &  0.06 &  0.31 \\
        & Fe &  0.26 &     5 &  0.06 &  0.09 &  0.13 &  0.04 &  0.15 &  0.04 &  0.23 \\
        \hline
        \hline\noalign{\vskip3pt}
  \end{tabular}}
  }\label{tab:result}
  \begin{tabnote}
  The calculation of $\sigma_\mathrm{SEM}$ is depicted in (1) of Section \ref{sec:error_estimate}. Those of $\sigma_{T_{\mathrm{eff}}}$,  $\sigma_{\log{g}}$, and $\sigma_{\xi}$ are described in (2). Those of $\sigma_\mathrm{OE}$  and  $\sigma_\mathrm{cont}$ are described in (3) and (4), respectively. $\sigma_\mathrm{Total}$ is the quadrature sum of all errors.
  \end{tabnote}
\end{table*}

We present the abundances obtained for each object in Table \ref{tab:result} with the contributions of individual error sources.

\subsection{Contributions of individual error sources} \label{sec:error_result}
We found that the error source that dominates the total error budget varies depending on the elements and objects.
In most cases, $\sigma_\mathrm{SEM}$, $\sigma_\mathrm{OE}$, and/or $\sigma_{\log{g}}$ are dominant, whereas $\sigma_{T_{\mathrm{eff}}}$ and $\sigma_\mathrm{cont}$ have minor contributions. %
This is different from the case of the chemical analysis of FGK dwarfs, in which the uncertainty of $T_{\mathrm{eff}}$ has a large impact on the derived elemental abundances.
The fact that $\sigma_\mathrm{OE}$ significantly contributes to the total error demonstrates the sensitivity of the line strengths to the abundance of elements other than the species responsible for the absorption, which is a phenomenon unique to M dwarfs (see Section \ref{sec:insensitivity}).
$\sigma_{\xi}$ has negligible contributions through all the elements of our objects.
The line broadening is controlled by other mechanisms in the temperature regime of M dwarfs.
One of these is pressure broadening, the treatment of which is described in Section \ref{sec:spectral_lines}.
The pressure broadening is subject to $\log{g}$, which is a reason for the large values of $\sigma_{\log{g}}$.
Note that the change in $\log{g}$ changes the ionization rate of some elements, which also changes the abundance results from neutral lines in the same direction as the effect of pressure broadening. %

For the two early-M dwarfs with temperatures $\sim$3700 K, $\sigma_\mathrm{SEM}$ and $\sigma_\mathrm{OE}$ are dominant %
for most elements, and $\sigma_{\log{g}}$ comparably contributes to the total errors for certain elements (Mg, K, Ca, and Ti). %
The relatively small $\sigma_{\log{g}}$ of HD$\:$233153 is attributed to the relatively small uncertainty of $\log{g}$ for the object.  %

For the three mid-M dwarfs with temperatures of $\sim$3200 K, the principal error source is $\sigma_\mathrm{OE}$ or $\sigma_{\log{g}}$ for many elements.
The absolute sizes of $\sigma_\mathrm{OE}$ are noticeably larger in the mid-M dwarfs (0.08--0.31) than in the early-M dwarfs (0.02--0.16).
The cause of this result is discussed in Section \ref{sec:insensitivity}. %
The absolute sizes of $\sigma_{\log{g}}$ are also larger in the mid-M dwarfs, and their trend found between different elements %
is almost the same as that found in the early-M dwarfs except for Ti, for which $\sigma_{\log{g}}$ is more significant in the mid-M dwarfs. %
This is partly because the pressure broadening becomes dominant in the Ti I line profiles as the temperature drops.
The main reason, however, is that the change in assumed $\log{g}$ alters the abundances of other elements assumed in the analysis, %
which causes a substantial change in the final Ti abundances (see Section \ref{sec:titanium}). %
The total errors of [Ti/H] for all the three mid-M dwarfs are larger than 0.3 dex.
We discuss the difficulty of constraining [Ti/H] in more detail in Section \ref{sec:titanium}.
$\sigma_{T_{\mathrm{eff}}}$ contributes to the total errors as much as the aforementioned two sources for certain elements, especially Na, K, and Ti. %
$\sigma_\mathrm{SEM}$ is also a main contributor to the total errors for many elements of G$\:$102-4 and $\rho^{01}\,$Cnc\,B. %
$\sigma_\mathrm{cont}$ is almost negligible in all elements except for Cr, where it is comparable to or larger than the errors of other sources, e.g., $\sigma_{\log{g}}$.  %

\subsection{Consistency check with G/K-type primaries} \label{sec:consistency_with_primaries} %

\begin{figure}
 \begin{center}
  \includegraphics[width=78mm]{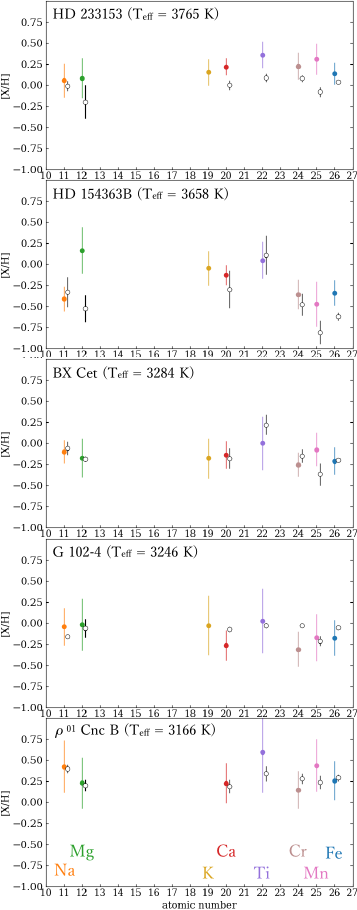}%
 \end{center}
 \caption{Abundance ratios [X/H] of individual elements for each binary pair as a function of atomic number.
 The color-filled circles represent the abundances of the M-dwarf secondaries we determined, and the black open circles (horizontally shifted for clarity) represent those of the G/K-type primaries reported by \citet{2018MNRAS.479.1332M}.
 The error bars for M dwarfs correspond to $\sigma_\mathrm{Total}$ in Table \ref{tab:result}, whereas those for the G/K-type primaries are the uncertainties reported by \citet{2018MNRAS.479.1332M}, which are calculated similarly to the $\sigma_\mathrm{SEM}$ in our analysis.
 }\label{fig:res_all}  %
\end{figure}

\begin{table}
  \tbl{Error-weighted RMS of the abundance differences of the M dwarfs from their G/K-type primaries for each element or object
  }{
  \begin{tabular}{lc}
    \hline\noalign{\vskip2pt}
        Species or object & Error-weighted RMS of differences \\[2pt]%
    \hline\noalign{\vskip2pt}
Na & 0.07 \\%
Mg & 0.33 \\%
Ca & 0.17 \\%
Ti & 0.21 \\%
Cr & 0.17 \\%
Mn & 0.29 \\%
Fe & 0.15 \\%
\hline\noalign{\vskip2pt}
HD$\:$233153 & 0.23 \\%
HD$\:$154363B & 0.30 \\%
BX$\:$Cet & 0.13 \\%
G$\:$102-4 & 0.16 \\%
$\rho^{01}\,$Cnc\,B & 0.12 \\%
    \noalign{\vskip2pt} \hline
    \end{tabular}} \label{tab:binary-check}
  \begin{tabnote}
  Potassium is excluded because its abundances for the primary stars were not reported. %
  \end{tabnote}
\end{table}

Figure \ref{fig:res_all} shows the comparison of chemical abundances between our results for the M-dwarf secondaries and the values for the G/K-type primaries reported by \citet{2018MNRAS.479.1332M} (hereafter Mon18), who determined the abundances using high-resolution visible spectra. %
The K abundances of our results are plotted here, to compare the abundance ratios to those of other elements, although the abundance of this element for the primaries was not determined by Mon18.
The abundances of the other seven elements determined for M dwarfs generally agree with those of the primaries within the error margins.
Note that the error bars for the abundances of the primaries reported by Mon18 contain only the line-to-line scatters divided by the square root of the number of spectral lines used in the analysis.
Mon18 specifically compiled the abundances of FGK stars in binary or multiple systems with M dwarfs.
We chose Mon18 for comparison because it was the most recent work to cover all the primary stars of the binary systems studied here.
Mon18 reported that their results agreed well with previously published ones %
and no significant offset was detected.

To examine the consistency between our resulting abundances of M dwarfs and the reported abundances of their primaries, %
we calculated the RMS of the abundance differences between an M dwarf and its primary for each element (five binary pairs) and for each object (seven elements).
In the process to obtain the mean, we gave a weight to each binary pair computed from the inverse of the quadrature sum of errors for both binary components.
The calculated values are given in Table \ref{tab:binary-check}.
The derived values are similar to or slightly smaller than the sizes of $\sigma_\mathrm{Total}$ in Table \ref{tab:result}, typically ranging from 0.15 to 0.25.
The exceptions are those for Mg, Mn, and HD$\:$154363B, for which the reasons are examined below. %

As shown in Fig. \ref{fig:res_all}, the large difference in abundance of Mg is solely due to the large difference seen in HD$\:$154363B (see Section \ref{sec:HD 154363B}).
The abundances of Mn, in contrast, appear to be systematically higher in M dwarfs than in the primaries.
Note that the obtained Mn abundance is even higher if we analyze the Mn I lines without considering the hfs.
Mon18 mentioned that the general trend of their [Mn/Fe] is not consistent with the Galactic chemical evolution trends studied for 1111 FGK stars by \citet{2012A&A...545A..32A}, possibly because of the scarcity of useful lines. %
From Figure A2 of Mon18, the difference is typically approximately 0.1 dex.
Assuming the actual [Mn/H] values of the primaries are higher than their estimates by 0.1 dex, the departure of our results from those of Mon18 decreases.
The large RMS value of the difference in abundance obtained for HD$\:$154363B is caused by Mg and Fe, as discussed in Section \ref{sec:HD 154363B}.

Finally, we calculated the error-weighted RMS of all the differences in abundance of each element for each binary pair (i.e., 35 data) to obtain 0.20 dex.
It was consistent with the typical estimated size of $\sigma_\mathrm{Total}$. %
In conclusion, assuming that the abundance measurements for GK stars are robust, our determination of the elemental abundances of early- to mid-M dwarfs is reliable within a typical precision of 0.2 dex. %

\section{Discussion} \label{sec:discussion}

\subsection{Sensitivity of line strength to abundances of elements other than the absorber} \label{sec:insensitivity}

\begin{figure}
 \begin{center}
  \includegraphics[width=80mm]{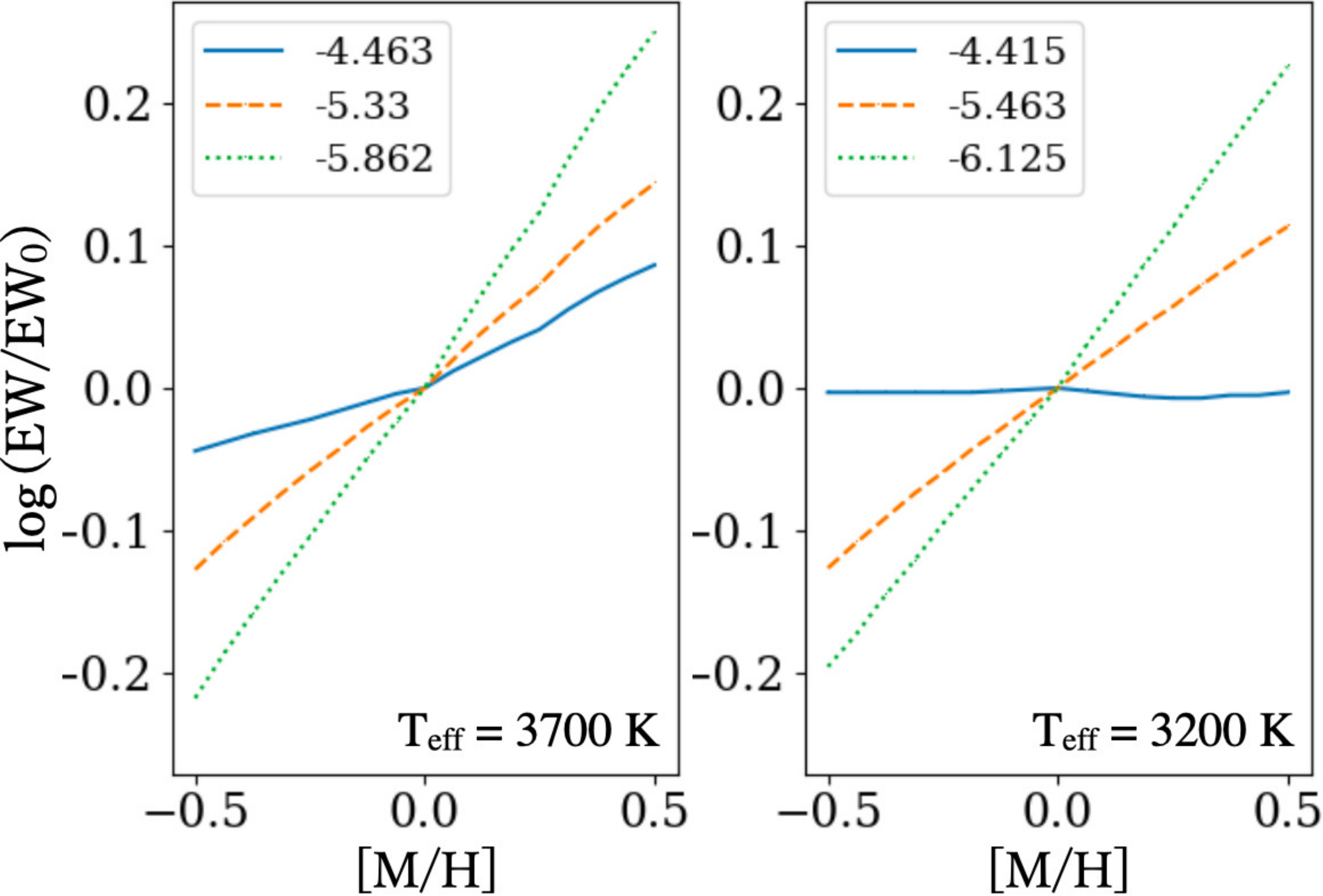}
 \end{center}
 \caption{Sensitivity of the EWs of Fe I lines in model spectra to the assumed overall metallicity [M/H]. The values on the vertical axis show the logarithmic EW normalized by the value at [M/H] = 0.0 ($\mathrm{EW}_0$). %
 $T_{\mathrm{eff}}$ is assumed to be 3700 K and 3200 K for the left and right panels, respectively. $\log{g}$ = 4.9 and $\xi$ = 0.5 for both panels.
 The blue solid line, orange dashed line, and green dotted line refer to relatively strong, medium, and weak lines, respectively. The legend in the upper-left corner shows $\log (\mathrm{EW}_0/\lambda)$, i.e., the logarithmic EW normalized by the wavelength in the case of [M/H] = 0.0.
  }\label{fig:ew_sensitivity_on_zlog}
\end{figure}

Through line selection for abundance analysis, we found that many atomic lines are not sensitive to variations in the overall metallicity, %
especially in the spectra of mid-M dwarfs.
Figure \ref{fig:ew_sensitivity_on_zlog} shows how the variation of overall metallicity assumed in the calculation alters the EWs of three Fe I lines in the model spectra as examples. 
It is widely recognized that stronger lines are less sensitive to absorber abundance due to the effect of saturation, as described by the so-called curve of growth. %
This explains the difference between the slopes of the three lines in both panels of Fig. \ref{fig:ew_sensitivity_on_zlog}.
However, the blue solid line in the right panel %
is almost flat, indicating that this strong line of mid-M dwarfs retains almost no sensitivity to the overall metallicity.
This cannot be explained by the saturation alone, and requires further explanation. %

\begin{figure}
 \begin{center}
  \includegraphics[width=80mm]{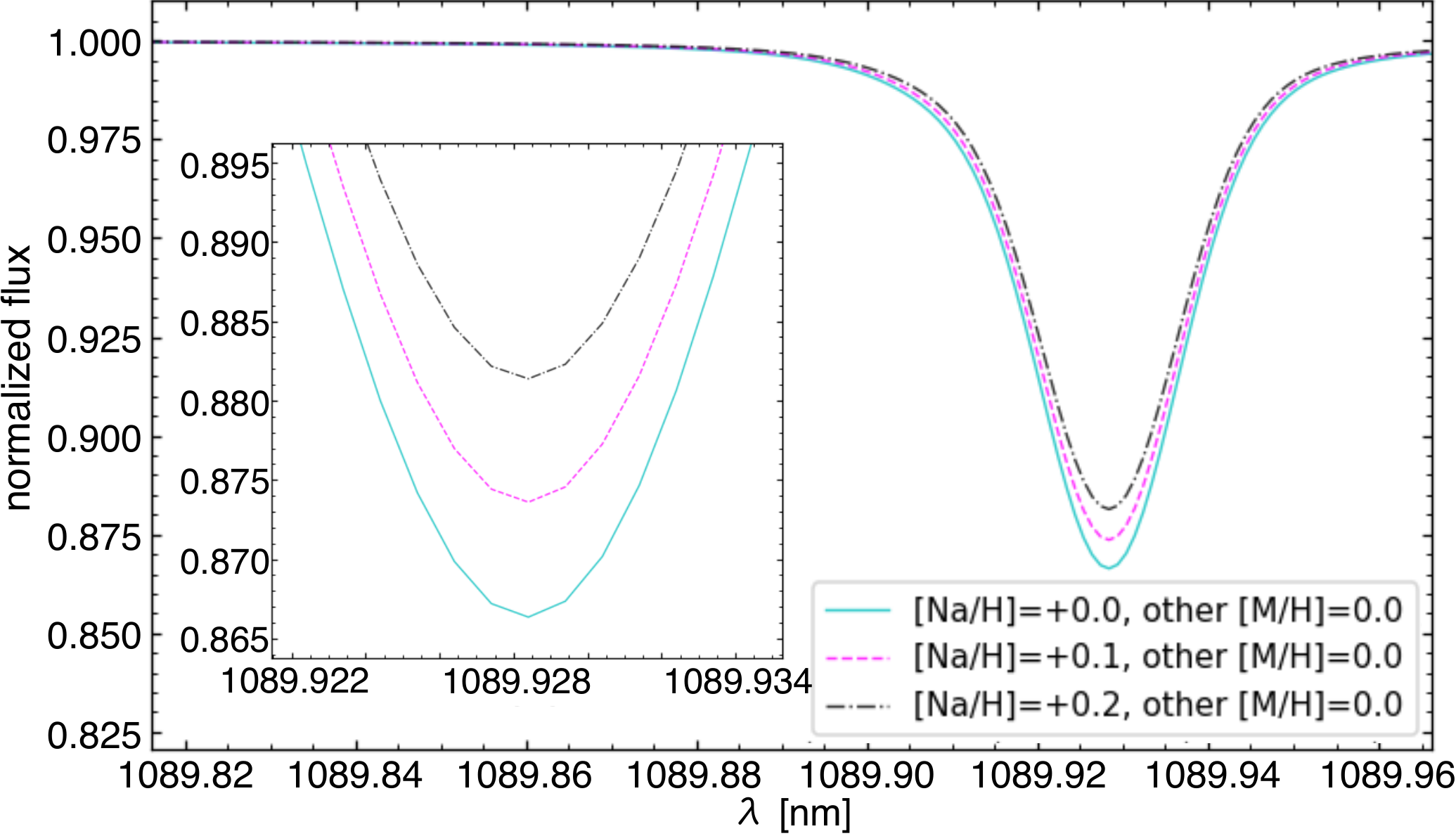}
 \end{center}
 \caption{Effect of Na abundance on an Fe I line. The three spectra were calculated assuming different Na abundances and the same solar abundances of other elements, including Fe. The internal panel presents an enlarged view of the line center.
 }\label{fig:effect_of_Na}
\end{figure}

We found that %
the low sensitivity of the absorption lines to the overall metallicity is attributed to the influence of elements other than the species responsible for the absorption.
For example, when higher metallicity is assumed, the higher Fe abundance strengthens the Fe I lines, but simultaneously, the higher abundances of other elements makes them weaker. %
In particular, elements with low ionization potential (e.g., Na, Ca) cause prominent effects.
The absorption profiles of an Fe I line with varying Na abundance assumed in the calculation are shown in Fig. \ref{fig:effect_of_Na}.
This demonstrates that an increase in Na abundance weakens the Fe I line.
Similar effects are also found when another element is increased. %

This phenomenon can be interpreted by considering the continuous absorption induced by negative hydrogen ions (H$^{-}$).
The bound-free transition of H$^{-}$ is the dominant opacity source at near-infrared wavelengths ($< 1.6  \, \mathrm{\mu m}$) in the atmospheres of stars similar to or cooler than the Sun. %
The increase in metal species, which are more or less ionized to release electrons, increases the electron pressure, promoting the binding reaction of hydrogen and electrons.
The resulting increase in H$^{-}$ increases the continuous absorption coefficient $\mathrm{\kappa_{\nu}}$. %
The larger $\mathrm{\kappa_{\nu}}$ reduces the continuum-normalized depths of absorption lines
 following the formula for the case of weak absorption lines (\citealt{2005oasp.book.....G}): %
\begin{equation}
\frac{F_{\mathrm{c}} - F_\nu}{F_{\mathrm{c}}} \sim \mathrm{constant} \, \frac{l_\nu}{\mathrm{\kappa_{\nu}}},
\label{eq:normalized depth}
\end{equation}
where $F_\nu$ is the flux at a certain wavelength $\nu$ inside the absorption line, $F_{\mathrm{c}}$ is the flux at wavelengths outside the absorption line, and $l_\nu$ is the line absorption coefficient at $\nu$.

\begin{figure}
 \begin{center}
  \includegraphics[width=80mm]{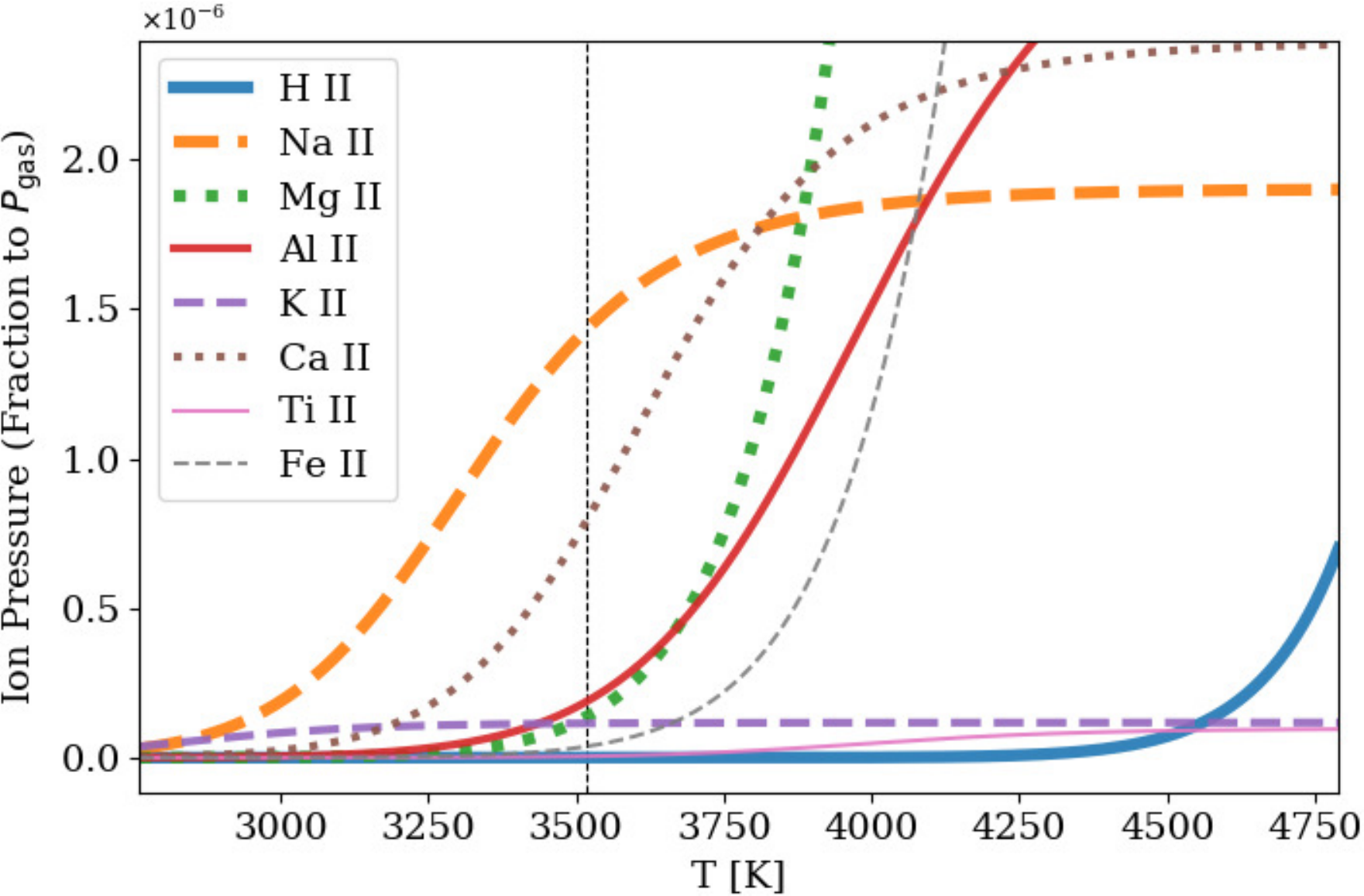}%
 \end{center}
 \caption{Degree of contribution of each element to the electron pressure as a function of temperature. %
 The vertical axis shows the ratio of the partial pressure of the singly ionized species to the total gas pressure. It is calculated assuming
 the total gas pressure $P_\mathrm{gas} = 1.2 \times 10^{6} \ \mathrm{dyn \ cm^{-2}}$,
 and
 the total electron pressure $P_\mathrm{e} = 3.4 \ \mathrm{dyn \ cm^{-2}}$,
 which are the values in the atmospheric depth with $\tau_{\mathrm{1200nm}} = 1.0$ under the condition of mid-M dwarfs with solar composition ($T_{\mathrm{eff}}$, $\log{g}$, $\xi$, $\mathrm{[M/H]}$) = (3200, 5.0, 0.0, 0.0).
 The atmospheric temperature at that depth is indicated by the vertical dashed line ($T_{\mathrm{eff}} = 3518$ K).
 Note that
 the values on the vertical axis presented for other temperatures do not necessarily match the actual fractions at the depth of the temperatures in the stellar atmosphere
 because of the fixed $P_\mathrm{gas}$ and $P_\mathrm{e}$.
 }\label{fig:saha}
\end{figure}

The effect of the change in abundance of a certain element on the continuous opacity depends on both its abundance and the ionization rate.
Atomic species with lower ionization potential and higher abundance have larger contributions to the electron pressure.
To elucidate which species dominate the effects, we solved Saha’s equation of ionization to determine the contribution of each element to the electron pressure.
The relative contributions of major species to the electron pressure are illustrated in Fig. \ref{fig:saha}.
We found that Na, Ca, and Mg are the dominant electron donors in the temperature range of early- and mid-M dwarfs, with many other species also contributing to some degree. %

The phenomena reported here also explain some features of the error sizes presented in Table \ref{tab:result}.
The $\sigma_\mathrm{OE}$ of an element becomes larger in mid-M than early-M dwarfs because the continuous opacity becomes more sensitive to the abundances of other elements. %
The small $\sigma_\mathrm{OE}$ of Na compared to other elements is attributed to the fact that the strength of Na I lines is relatively insensitive to the change in abundance of other elements,
because Na itself is the most dominant electron source. The Na abundance controls continuous opacity to a greater extent than that of the other elements.
For mid-M dwarfs, the $\sigma_{\log{g}}$ values are large even for elements for which the lines do not have broad wings. %
This is due to the sensitivity of the abundance results of Na, Mg, and Ca, which are obtained from broad lines, to the change in $\log{g}$.
In the iterative procedure, the variations in these abundance results also induce a substantial change in the abundance results of the other elements through the change in continuous opacity. %

In conclusion, %
we argue that it is risky to independently determine the abundance ratios of certain elements without considering the variation in abundance of other elements. %
Consistent abundance ratios for each object need to be applied to the analysis to obtain abundance ratios and even the overall metallicity.

\subsection{Titanium lines} \label{sec:titanium}
It is difficult to constrain the Ti abundance despite the large number of notable Ti I lines in the near-infrared spectra of M dwarfs.
The Ti I lines show little response to the overall metallicity in the spectra of early-M dwarfs. A negative correlation even appears between the Ti I line strengths and the overall metallicity at $T_{\mathrm{eff}} < $ 3400 K. %
This is a unique behavior found only for the absorption lines of Ti among all the elemental lines used.

\begin{figure}
 \begin{center}
  \includegraphics[width=80mm]{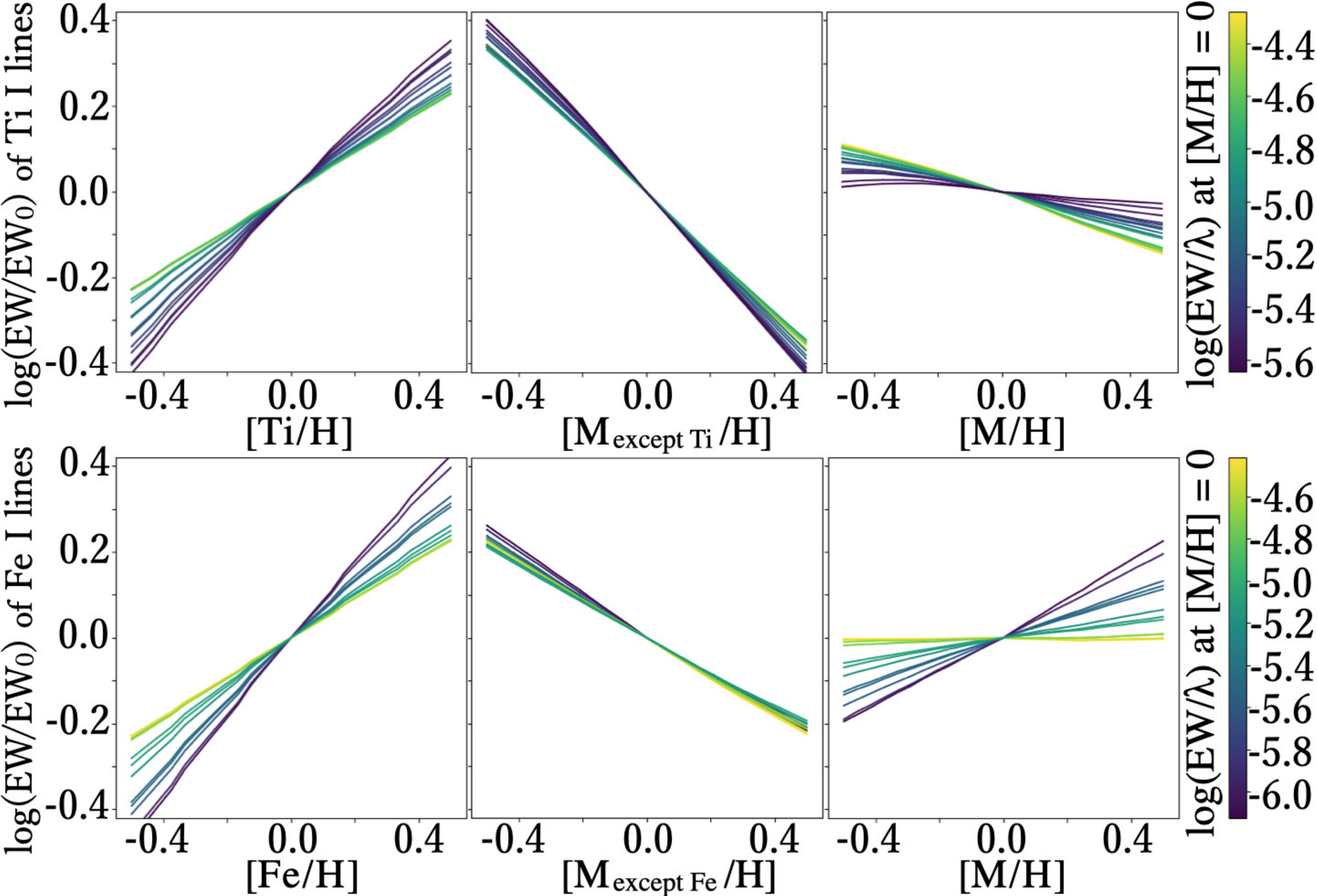}
 \end{center}
 \caption{Upper panels show the sensitivity of the EWs of all Ti I lines used in the analysis to the Ti abundance in the left panel, to the abundances of all elements except for Ti in the center panel, and to the overall metallicity including Ti in the right panel. %
 The vertical axis is the same as that in Fig. \ref{fig:ew_sensitivity_on_zlog}.
 All the spectra were calculated with $T_{\mathrm{eff}}$ = 3200 K, $\log{g}$ = 4.9, and $\xi$ = 0.5 km s$^{-1}$.
 The color bar indicates $\log (\mathrm{EW}_0/\lambda)$, i.e., the logarithmic EW at [M/H] = 0.0, normalized by the wavelength. %
 The lower three panels are the same plots as in the upper panels but for Fe instead of Ti.
 }\label{fig:ti_inverse_sensitivity}
\end{figure}

This is attributed to the abundance changes of elements other than Ti. %
We investigated the response of the Ti I lines in the model spectra separately to the abundances of Ti and of all the other elements, as shown in the upper three panels in Fig. \ref{fig:ti_inverse_sensitivity}.
The left panel shows the natural behavior of Ti I lines getting stronger as the Ti abundance increases.
The weaker lines show higher sensitivity to Ti abundance, as expected from a curve of growth.
The middle panel, in contrast, shows that the increase in the other elemental abundances weaken the Ti I lines.
This relation does not have a significant dependence on individual absorption lines. %
The negative slope shown here is steeper than the positive slope of the left panel.
As a result of both effects, the EWs of the Ti I lines gradually decrease as the overall metallicity increases, as shown in the right panel.

The same plots for Fe are shown for comparison in the lower panels of Fig. \ref{fig:ti_inverse_sensitivity}. %
The value of ``slope’’ hereafter is simply calculated as the change of the $\log{\mathrm{(EW)}}$ corresponding to the change of [X/H] from $-$0.5 to $+$0.5.
The slopes in the upper left panel of Fig. \ref{fig:ti_inverse_sensitivity} for Ti range from 0.45 to 0.78, which are not appreciably different from the slopes in the lower left panel for Fe. %
In contrast, the slopes of the plots in the upper middle panel range from $-$0.82 to $-$0.68, which are significantly steeper than any plots of the other elements, including those for Fe in the lower middle panel. %
Hence, the unique behavior found for Ti I lines is not due to the sensitivity of the line strengths to [Ti/H], but rather the abundances of other elements.

\begin{figure}
 \begin{center}
  \includegraphics[width=60mm]{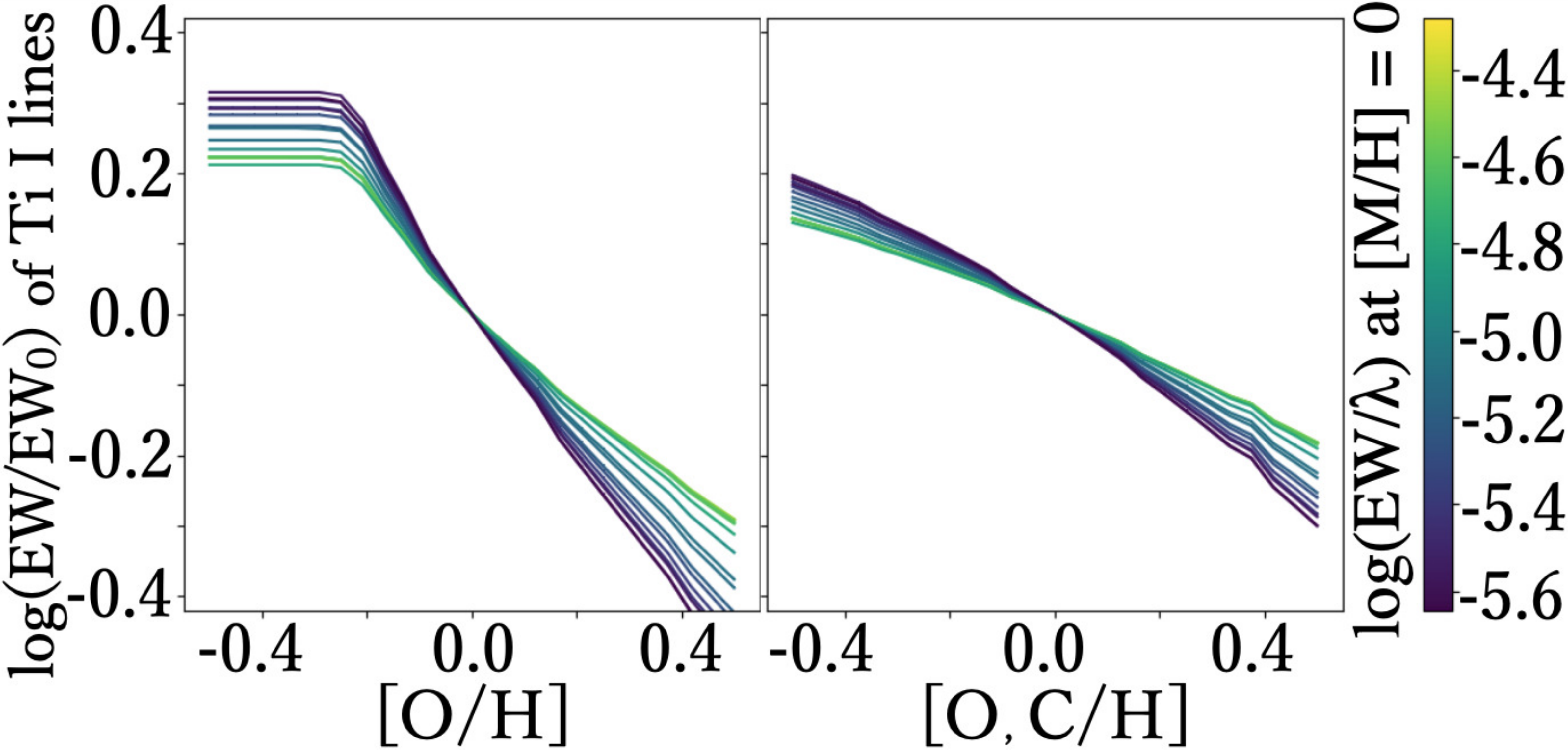}%
 \end{center}
 \caption{Same plots as in the upper left panel of Fig. \ref{fig:ti_inverse_sensitivity}, but for oxygen abundance [O/H] in the left panel, and for oxygen and carbon abundance [O, C/H] in the right panel.
  }\label{fig:oxygen_vs_Tilines}
\end{figure}

We found that the Ti I line strengths have a negative correlation with the oxygen abundances assumed in the calculation. %
Figure \ref{fig:oxygen_vs_Tilines} shows the modeled EWs of the Ti I lines as functions of [O/H] and [O, C/H]. The EWs sharply decrease with the increase in oxygen.
This is due to the decrease in neutral titanium (Ti I) as a result of the formation of titanium oxide molecules (TiO) in M dwarfs.
Note that the flat part appearing at [O/H] $\leq$ $-$0.25 dex in the left panel is due to the depletion of oxygen available to form TiO. %
The oxygen below this level is mainly combined with carbon to form CO; thus, the flat part is not seen if [C/H] also changes together with [O/H], as shown in the right panel.
The sensitivity of Ti I lines to O and C abundances assumed in the analysis explains the particularly steep slopes for the Ti I lines in the upper middle panel of Fig. \ref{fig:ti_inverse_sensitivity}.
We note that the influence of the variation in [O, C/H] on the strength of any other atomic lines used is negligible.

For a complete understanding of the peculiar behavior of Ti I lines, we also examined the influence of additional possible factors other than elemental abundances. %
Firstly, we investigated the response of Ti I lines to temperature.
Although the sensitivity of EW to $T_{\mathrm{eff}}$ differs from line to line depending on the $E_\mathrm{low}$ of the transition and the line strength,
the variation in EW corresponding to a 100 K change in $T_{\mathrm{eff}}$ is at most 0.025 dex. %
When the overall metallicity varies by 0.1 dex, the temperature of the atmospheric layer with $\tau \sim 1$ changes by 100 K because of the so-called line-blanketing effect, which can induce a 0.025 dex change in the EW of Ti I lines in an extreme case.
However, this variation is substantially smaller than the variation in Ti I lines due to changes in the oxygen abundances.

Secondly, we examined the influence of the nature of the absorption lines, i.e., the gf-value, $E_\mathrm{low}$, or EW.
To disentangle their separate effects and dependence on elemental species,
we probed the behavior of three artificial lines of Cr, the ionization energy (6.764 eV) of which is close to that of Ti (6.820 eV). %
We set up the parameters of each artificial line, namely the gf-value, $E_\mathrm{low}$, and EW, such that two were identical to those of the actual Ti I line at 967.8197 nm. %
As a result, the sensitivity of all the artificial Cr I lines to the variation in abundance of the other elements was shown to be similar to that of the actual Cr I lines, and less than that of the Ti I lines. %
This indicates that the unique behavior of Ti I lines is not due to any specific nature of the spectral lines.

We conclude that the dominant cause of the unique response of Ti I lines to changes in the overall metallicity is the consumption of neutral Ti atoms by the formation of TiO molecules. %
This behavior was briefly argued by \citet{1976ApJ...210..402M}, who found a similar negative metallicity dependence of Ti I lines in the $I$-band spectra of mid-M dwarfs and mentioned the effect of TiO formation. %
\citet{1973A&A....23..411T} demonstrated the partial pressure of each species harboring Ti in stellar atmospheres at different $T_{\mathrm{eff}}$, as shown in Fig. 3 of the paper.
The abundance of TiO exceeds that of neutral Ti at temperatures less than approximately 2800 K, while Ti$^+$ ions dominate over neutral Ti at temperatures higher than approximately 3200 K. %
Because the atmospheric temperature of M dwarfs corresponds to the temperature range in which these transitions occur, the changes in the ionization state of Ti complicate the interpretation of the Ti I lines. %
Similar transitions are not seen for the other elements analyzed in this study, whereas silicon and zirconium show similar behavior, forming SiO and ZrO at $T < $ 3000 K.
Special care is required for the analysis of Si I or Zr I lines. %

Several previous studies used Ti I lines to estimate the [M/H] or [Fe/H] of M dwarfs (e.g., \citealt{2019A&A...627A.161P}, \citealt{2018A&A...620A.180R}, \citealt{2016A&A...586A.100L}). %
However, the unique behavior of the lines has not been fully discussed.
Here, we urge caution in using the the Ti I lines as the indicator of [M/H], because the increases in Ti and in the other elements have opposite effects on the strength of the lines. %
When we derive [Ti/H], we need to analyze the Ti I lines with reliable abundances of other elements,
especially O, C, Na, and Ca.
\citet{2017ApJ...851...26V} and \citet{2018ApJ...863..166V} also used the Ti I lines as indicators of the Ti abundance and the stellar age.
The negative correlation between the Fe abundances and EWs of Ti I lines can be seen in some EW pairs measured and tabulated by \citet{2017ApJ...851...26V}.
However, they did not estimate the impact of other elements, such as O and Na, which could cause potential uncertainty in their estimates.

\subsection{HD 154363B} \label{sec:HD 154363B}
Here, we discuss the noticeable discrepancy in the Fe and Mg abundances between HD$\:$154363B and its primary HD$\:$154363A, as seen in Fig. \ref{fig:res_all}.
Our abundances of HD$\:$154363B ([Fe/H] = $-0.34 \pm 0.15$; [Mg/H] = $0.17 \pm 0.27$) deviates from the error range of those of the primary ([Fe/H] = $-0.62 \pm 0.05$; [Mg/H] = $-0.53 \pm 0.16$). %

\begin{figure}
 \begin{center}
  \includegraphics[width=80mm]{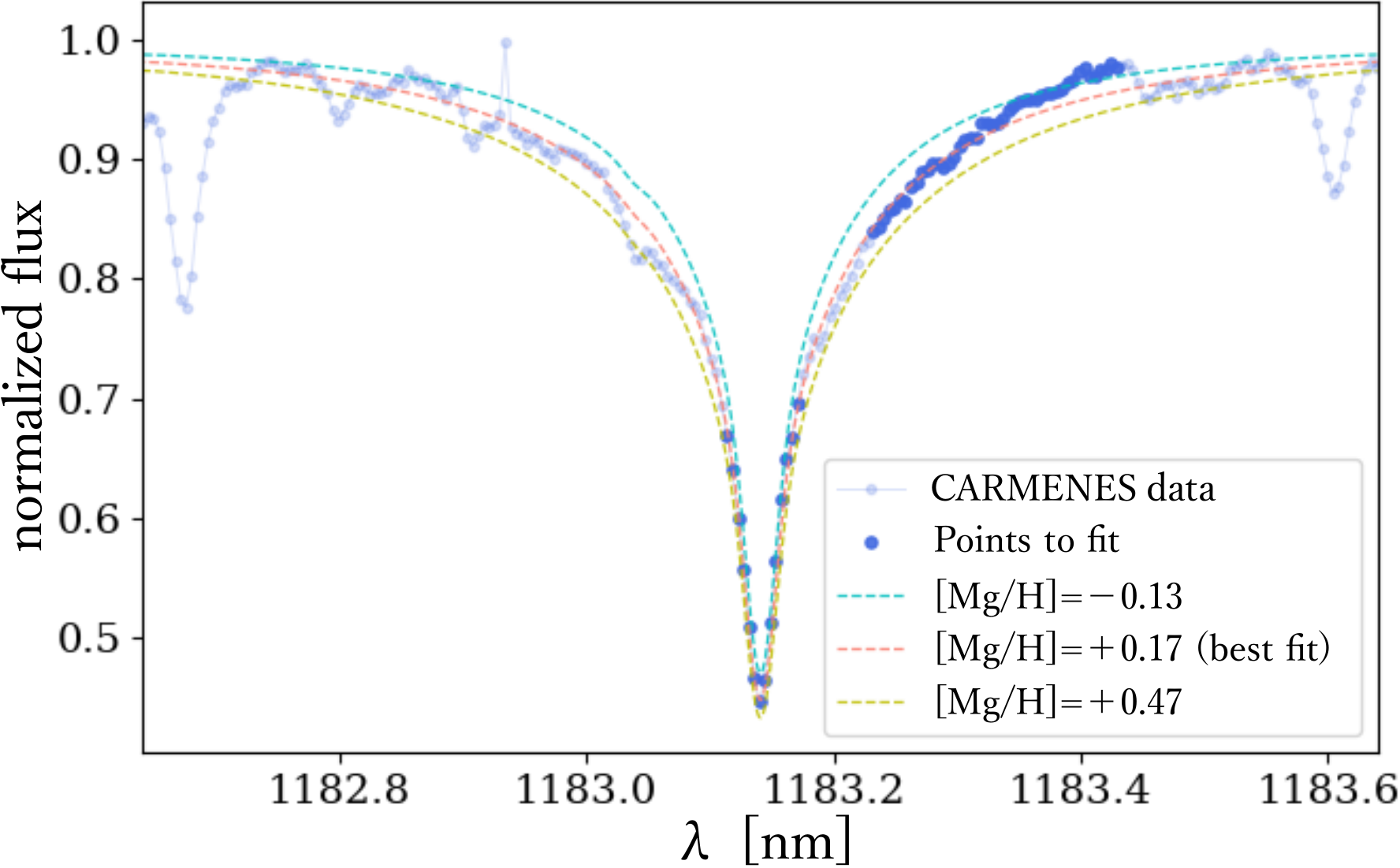}
 \end{center}
 \caption{Mg I line in the observed spectra of HD$\:$154363B (blue dots; the deep dots are used for fitting and the pale dots are not used) and synthetic model spectra calculated with the best-fit Mg abundance [Mg/H] = 0.13 dex, and with 0.3 dex higher and lower abundances (dashed lines). }\label{fig:Mg_line_shape}  %
\end{figure}

The Mg I line used is relatively deep and broad (EW $\sim$107 pm) with wide damping wings.
To examine the effect of possible imperfection of the Voigt profile fitting,
we conducted a chi-square fitting between the observed line profile and synthetic spectra (Fig. \ref{fig:Mg_line_shape}). %
As a result, we derived almost the same abundance result as the EW analysis, confirming that the [Mg/H] of this M dwarf is higher than the value obtained by Mon18 for the primary star. %

The overall metallicity of HD$\:$154363B has been studied based on empirical calibration methods. %
\citet{2014MNRAS.443.2561G} derived [Fe/H] $ = -0.39 \pm 0.11$ from the medium-resolution ($R \sim$ 1300) visible spectra.
\citet{2014AJ....147...20N} obtained [Fe/H] $ = -0.44 \pm 0.14$ from the medium-resolution ($R \sim$ 2000) near-infrared spectra.
Both are consistent with our Fe abundance of HD$\:$154363B rather than that of its primary as reported by Mon18.

Lastly, %
our results with low [Fe/H] and enhanced [Mg/Fe] ($0.51 \pm 0.31$) follow the abundance trends in the thick disk of the Milky Way Galaxy (e.g.,  \citealt{2014A&A...562A..71B}). %
Although the reason for the discrepancy between our result for HD154363B and that of Mon18 for the primary star is not clear, we regard our abundance results for this M dwarf as reliable from the above inspection. %

\section{Summary} \label{sec:summary}
We analyzed the high-resolution near-infrared CARMENES spectra of five M dwarfs belonging to binary systems with G or K dwarfs to determine their abundances of seven to eight individual elements. %
We verified our determination by comparing our abundances of M dwarfs with those of the primary stars derived by \citet{2018MNRAS.479.1332M} from high-resolution visible spectra.
The results show good agreement, typically within 0.2 dex, without significant systematics. %

We calculated the expected precision of the result for each element of each object from the quadrature sum of the errors arising from six different sources to be typically 0.1--0.3 dex.
We also investigated the contribution of individual sources to the total error.
The dominant error source depends on the elemental species and the stellar temperature.
A notable difference from the standard analysis for FGK stars is the large errors propagated from the uncertainties of $\log{g}$ and the abundances of elements other than the corresponding species.
The derived abundances have high sensitivity to $\log{g}$ because many lines used in the analysis are affected by pressure broadening under the high $\log{g}$ of M dwarfs.
The abundance results are also affected by the variation of the continuous opacity by H$^{-}$, which depends on the abundances of metals with low ionization potential, such as Na and Ca.
In addition, the strength of Ti I lines in the near-infrared spectra of M dwarfs with $T_{\mathrm{eff}} < $ 3400 K shows an anti-correlation with the overall metallicity %
 because the increase in oxygen promotes the consumption of neutral Ti by the formation of TiO molecules.

These results indicate that to constrain the true abundance of any element or even the overall metallicity, it is crucial to determine the consistent abundance ratios of individual elements in the chemical analysis of each M dwarf.

The application of our chemical analysis to the M dwarfs observed by the planet search projects provides reliable and consistent abundance ratios on a larger sample of nearby M dwarfs than ever before. %
This information will be useful for observational understanding of how the occurrence rate and characteristics of planets depend on the elemental abundances of the host M dwarfs. %

\begin{ack}
  We thank the anonymous referee for helpful and constructive comments. %
  This research is based on data from the CARMENES data archive at CAB (INTA-CSIC).
  This research is also based on observations collected at the Centro Astronómico Hispano-Alemán (CAHA) at Calar Alto, operated jointly by Junta de Andalucía and Consejo Superior de Investigaciones Científicas (IAA-CSIC). %
  Data analysis was in part carried out on the Multi-wavelength Data Analysis System operated by the Astronomy Data Center (ADC), National Astronomical Observatory of Japan. %
  This work made use of the VALD database, operated at Uppsala University, the Institute of Astronomy RAS in Moscow, and the University of Vienna. %
  We acknowledge the financial support from the Astrobiology Center, NINS, for purchasing the observing time of CARMENES in CAHA. %
  This work was partially supported by the SOKENDAI Long-term Internship Program and Overseas Travel Fund for Students of the Department of Astronomical Science, The Graduate University for Advanced Studies, SOKENDAI. %
  This research made use of Astropy, %
  a community-developed core Python package for Astronomy (Astropy Collaboration 2013, 2018). \nocite{2013A&A...558A..33A, 2018AJ....156..123A} %
  We would like to thank Editage (www.editage.com) for English language editing. %
\end{ack}

\section*{Supporting Information}
Table \ref{tab:EWs_excerpt} is available in its entirety in machine-readable format in the online version of this article.
All figures are in color online.

\bibliographystyle{apj} %
\bibliography{library} %

\appendix
\renewcommand{\thetable}{\Alph{section}.\arabic{table}}\setcounter{table}{0}
\renewcommand{\thefigure}{\Alph{section}.\arabic{figure}}\setcounter{figure}{0}

\section{Exclusion of BD-02$\;$2198} \label{sec:BD-02 2198} %

\begin{table*}[bthp]%
  \tbl{Basic information of BD-02$\;$2198 (same format as Table \ref{tab:targets})}{
  %\small
  \renewcommand{\arraystretch}{1.0}
  \scalebox{0.9}[1.0]{ %{0.87}[1.0]{ %
  \begin{tabular}{p{16mm}cccccp{18mm}ccc}
    \hline\noalign{\vskip3pt}
        \multicolumn{1}{c}{Name}  &  RA  &  Dec  &  SpT  &  $T_{\mathrm{eff}}$  &  $\log{g}$  &  Name$\;$of  &  SpT of  &  [Fe/H]$\;$of  &  S/N  \\ %
        &  \multicolumn{2}{c}{J2000.0}  && (K) && Primary  &  Primary  &  Primary  &  (1000 nm)  \\  [2pt]
    \hline\noalign{\vskip3pt}
BD-02$\;$2198  &  \timeform{07h36m07.08s}  &  \timeform{-03D06'38.8''}  &  M1.0V  &  3744$\pm$82  &  4.76$\pm$0.24  &  HD$\;$61606$\;$A,$\:$B  &  K3V,$\:$K7V  &  $-$0.11$\pm$0.03  &  187  \\
\hline\noalign{\vskip3pt}
\end{tabular}}}\label{tab:bd02_2198_targets}
\end{table*}

\begin{table*}[bthp]
  \tbl{Abundance results and errors (same format as Table \ref{tab:result})}{
  \begin{tabular}{llccccccccc}
    \hline\noalign{\vskip3pt}
        \multicolumn{1}{c}{Object} & Element &  [X/H]  &  $N_\mathrm{line}$  &  $\sigma_\mathrm{SEM}$  &  $\sigma_{T_{\mathrm{eff}}}$  &  $\sigma_{\log{g}}$  &
        $\sigma_{\xi}$  &  $\sigma_\mathrm{OE}$  &  $\sigma_\mathrm{cont}$  &  $\sigma_\mathrm{Total}$ \\
    \hline\hline\noalign{\vskip3pt}
        BD-02$\;$2198
      & Na & $-$0.04 &     2 &  0.08 &  0.01 &  0.01 &  0.00 &  0.02 &  0.05 &  0.10 \\
      & Mg &  0.29 &     1 &  0.12 &  0.09 &  0.05 &  0.02 &  0.09 &  0.05 &  0.19 \\
      & K &  0.44 &     2 &  0.08 &  0.02 &  0.08 &  0.01 &  0.06 &  0.03 &  0.14 \\
      & Ca &  0.32 &     6 &  0.07 &  0.01 &  0.04 &  0.01 &  0.05 &  0.03 &  0.10 \\
      & Ti &  0.48 &    10 &  0.04 &  0.05 &  0.03 &  0.03 &  0.15 &  0.02 &  0.17 \\
      & Cr &  0.42 &     3 &  0.07 &  0.05 &  0.01 &  0.03 &  0.08 &  0.05 &  0.13 \\
      & Mn &  0.17 &     1 &  0.12 &  0.05 &  0.02 &  0.02 &  0.10 &  0.05 &  0.18 \\
      & Fe &  0.28 &     5 &  0.05 &  0.05 &  0.02 &  0.03 &  0.08 &  0.05 &  0.13 \\
      \hline
\hline%
\end{tabular}}
\label{tab:bd02_2198_result}
\end{table*}

In addition to the sample given in Table 1, we analyzed the spectra of another early-M dwarf, BD-02$\;$2198, obtained from the CARMENES GTO Data Archive.
The basic information and abundance results are presented in Tables A1 and A2, respectively. %
However, we excluded it from our targets in this paper,
based on the kinematic information. %

\citet{2009ApJ...706..343P} reported that this star is a member of a triplet with HD$\;$61606$\;$A and B based on their proper motions, parallaxes, radial velocities, and ages estimated by the X-ray luminosity.
However, they also reported that the difference in their proper motions is larger than the error, and suggested the possibility of dynamical disintegration.
We used the latest astrometric data from $Gaia$ DR2 (\citealt{2018A&A...616A...1G}) to compare the escape velocity and actual three-dimensional relative velocity of BD-02$\;$2198 with respect to the barycenter of the HD$\;$61606$\;$AB system.
Here, we treated the two K dwarfs as one mass point for simplicity because their distance from BD-02$\;$2198 is almost an order of magnitude larger than the distance between them.
The mass of HD$\;$61606$\;$A ($0.82 \pm 0.03 \; M_\odot$) is sourced from \citet{2017AJ....153...21L} and those of HD61606B ($0.66 \pm 0.07 \; M_\odot$) and BD-02$\;$2198 ($0.54 \pm 0.07 \; M_\odot$) are sourced from \citet{2014MNRAS.443.2561G}.
The escape velocity of BD-02$\;$2198 from HD$\;$61606$\;$AB is %
\begin{equation}
v_{\mathrm{escape}} = \sqrt{\frac{2 \mathrm{G} (M + m)}{r}},
\label{eq:escape velocity}
\end{equation}
where G is the gravitational constant, $M$ is the total mass of HD$\;$61606$\;$AB, $m$ is the mass of BD-02$\;$2198, and $r$ is the distance of BD-02$\;$2198 from the barycenter of HD$\;$61606$\;$AB.
The result is $0.25 \pm 0.01$ km s$^{-1}$.
The relative velocity of BD-02$\;$2198 to HD$\;$61606$\;$AB estimated from their proper motions and radial velocities is $1.58 \pm 0.43$ km s$^{-1}$, which exceeds the escape velocity by more than 3$\sigma$. %
This confirms that BD-02$\;$2198 is not gravitationally bound by HD$\;$61606$\;$AB. %

\begin{figure}%
 \begin{center}
  \includegraphics[width=78mm]{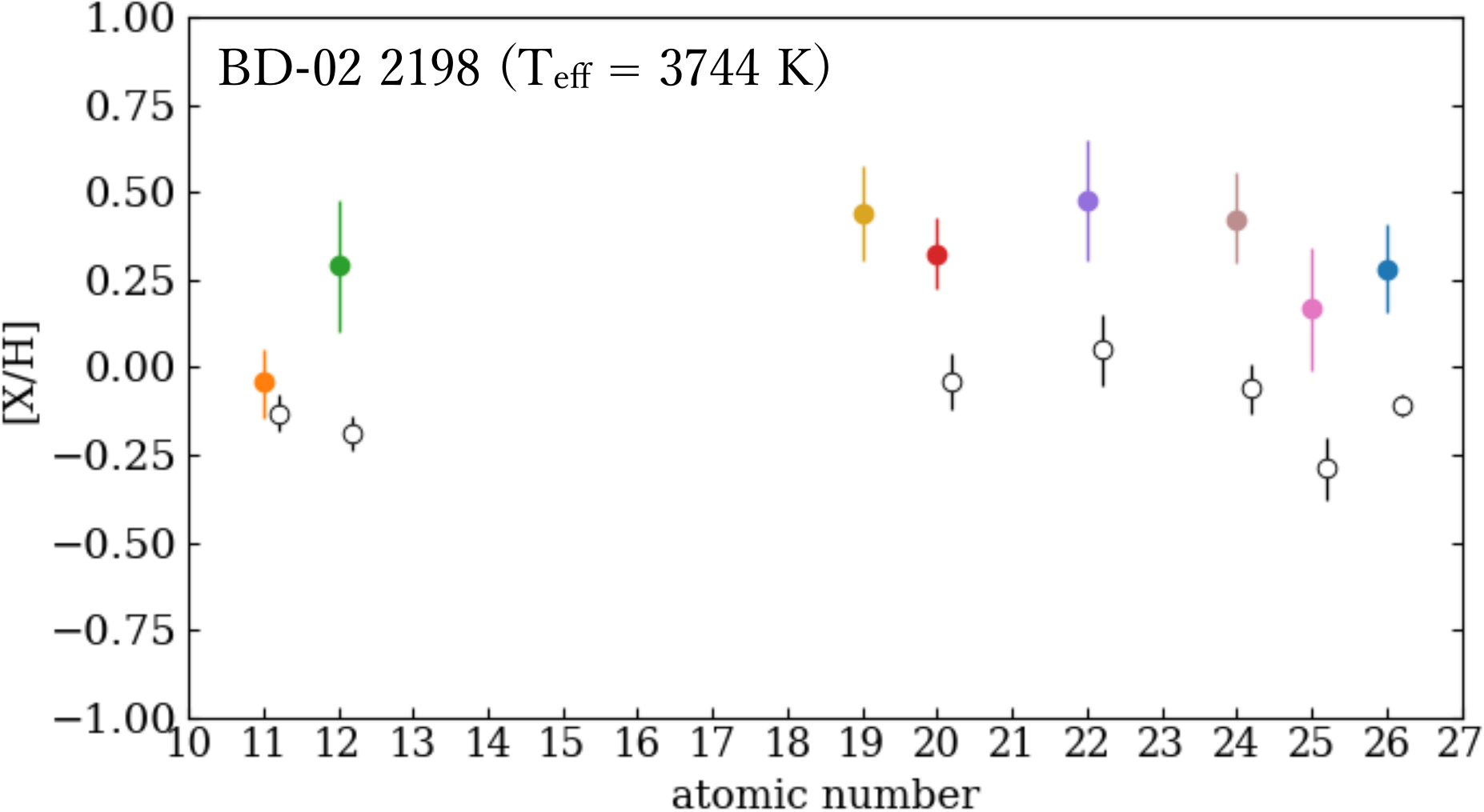}%
 \end{center}
 \caption{Abundances of individual elements we determined for BD-02$\;$2198 (color-filled circles), and those for HD$\;$61606$\;$A determined by \citet{2018MNRAS.479.1332M} (black open circles).
 The axes are the same as in Fig. \ref{fig:res_all}.
 }\label{fig:bd02_2198}
\end{figure}

The results of our abundance analysis on BD-02$\;$2198 show an offset greater than the error margin of the abundances of HD$\;$61606$\;$A reported by Mon18, as seen in Fig. \ref{fig:bd02_2198}. %
We attributed the discrepancy not to the systematic overestimation of our analysis but to the different origins of these objects.
Based on the above investigation, we concluded that BD-02$\;$2198 and HD$\;$61606$\;$AB are not members of the same multiple system sharing a common origin,
and thus excluded them from the binary comparison in Section \ref{sec:consistency_with_primaries}.

\end{document}